\documentclass[12pt]{article}
\usepackage{ametsoc}
\usepackage{amsmath} 
\usepackage{amssymb} 
\usepackage{amsfonts}
\usepackage{graphicx,color}                 

\newcommand{\myabstract}{
We propose a robust ensemble filtering scheme based on the $H_{\infty}$ filtering theory. The optimal $H_{\infty}$ filter is derived by minimizing the supremum (or maximum) of a predefined cost function, a criterion different from the minimum variance used in the Kalman filter. By design, the $H_{\infty}$ filter is more robust than the Kalman filter, in the sense that the estimation error in the $H_{\infty}$ filter in general has a finite growth rate with respect to the uncertainties in assimilation, except for a special case that corresponds to the Kalman filter.

The original form of the $H_{\infty}$ filter contains global constraints in time, which may be inconvenient for sequential data assimilation problems. Therefore we introduce a variant that solves some time-local constraints instead, and hence we call it the time-local $H_{\infty}$ filter (TLHF). By analogy to the ensemble Kalman filter (EnKF), we also propose the concept of ensemble time-local $H_{\infty}$ filter (EnTLHF). We outline the general form of the EnTLHF, and discuss some of its special cases. In particular, we show that an EnKF with certain covariance inflation is essentially an EnTLHF. In this sense, the EnTLHF provides a general framework for conducting covariance inflation in the EnKF-based methods. We use some numerical examples to assess the relative robustness of the TLHF/EnTLHF in comparison with the corresponding KF/EnKF method.
}
\begin{document}
%
%
\title{\textbf{\large{Robust ensemble filtering and its relation to covariance inflation in the ensemble Kalman filter}}}
%
%
\author{\textsc{Xiaodong Luo}
				\thanks{\textit{Corresponding author address:}
				4700 King Abdullah University of Science and Technology, Thuwal, 23955-6900, Saudi Arabia
				\newline{E-mail: xiaodong.luo@kaust.edu.sa}}\quad\textsc{and Ibrahim Hoteit}\\
\textit{\footnotesize{King Abdullah University of Science and Technology, Thuwal, Saudi Arabia}}
}
%
\ifthenelse{\boolean{dc}}
{
\twocolumn[
\begin{@twocolumnfalse}
\amstitle

\begin{center}
\begin{minipage}{13.0cm}
\begin{abstract}
	\myabstract
	\newline
	\begin{center}
		\rule{38mm}{0.2mm}
	\end{center}
\end{abstract}
\end{minipage}
\end{center}
\end{@twocolumnfalse}
]
}
{
\amstitle
\begin{abstract}
\myabstract
\end{abstract}
}

\section{Introduction}

The Kalman filter (KF) \citep{Kalman-new} is a sequential data assimilation algorithm. For linear stochastic systems, it can be shown that the KF is an optimal linear estimator that minimizes the variance of the estimation error \cite[ch.~5]{Simon2006}. Because of its relative simplicity in implementation, the KF is suitable for many data assimilation problems. However, for high dimensional systems such as weather forecasting models, direct application of the KF is prohibitively expensive as it involves manipulating covariance matrices of the system states. For this reason, different modifications of the KF were proposed to reduce the computational cost. These include various ensemble Kalman filters (EnKF) \citep{Anderson-ensemble,Bishop-adaptive,Burgers-analysis,Evensen-sequential,Evensen-assimilation,Houtekamer1998,Whitaker-ensemble}, the error subspace-based filters \citep{Cohn1996-approx,Hoteit2001,Hoteit2002,Luo-ensemble,Pham1998,Verlaan1997}, and filters based on other strategies \citep{Beezley-morphing,Zupanski-maximum}, to name but a few. A detailed description of the above filters is beyond the scope of this work. Readers are referred to \cite{Evensen-ensemble,Nerger2005,Tippett-ensemble} for reviews of some of the aforementioned filters. Roughly speaking, these modifications exploit the information of a subset in the state space of a dynamical system, while the information of the complement set is considered less influential, and thus ignored. Consequently, the computations of these modified filters are normally conducted on the chosen subsets, instead of the whole state space, so that their computational costs are reduced. For simplicity, we may sometimes abuse the terminology by referring to all the aforementioned filters as the EnKF-based methods (EnKF methods for short).

The KF and the EnKF are among the family of Bayesian filters that adopt Bayes' rule to update background statistics to their analysis counterparts. In these filters, one needs to make certain assumptions on the statistical properties (e.g., pdfs or moments) of both the dynamical and observation systems. In reality, however, these assumptions may not be accurate, so that a Bayesian filter may fail to achieve good performance with mis-specified statistical information \citep{Schlee1967}. For example, if implemented straightforwardly, an EnKF with a relatively small ensemble size may produce inaccurate estimations of covariance matrices \citep{Whitaker-ensemble}. This could degrade filter performance, or even cause filter divergence. As a remedy, in practice it is customary to conduct covariance inflation and localization to relieve this problem \citep{Anderson-Monte,Hamill-distance,Hamill2009,VanLeeuwen2009}.

In contrast, robust filters emphasize on the robustness of their estimates, so that they may have better tolerances to possible uncertainties in assimilation. The estimation strategies of robust filters are different from Bayes' rule. Take the $H_{\infty}$ filter \citep{Francis1987,Simon2006}, one of the robust filters, as an example. The $H_{\infty}$ filter (HF) does not assume to exactly know the statistical properties of a system in assimilation. Instead, it accepts the possibility that one may only have incomplete information of the system. Consequently, rather than looking for the best possible estimates based on Bayes' rule, the optimal $H_{\infty}$ filter employs a robust strategy, namely, the minimax rule \citep[ch.~4]{Burger1985}, to update its background statistics. This robustness may be of interest in practical situations. For example, for data assimilation in earth systems, the system models are often not the exact descriptions of the underlying physical processes, and it is challenging to characterize the properties of the corresponding model errors (\citealp{Wang2008455} and the references therein). Given an imperfect model, the estimation error of the HF in general grows with the uncertainties in assimilation at a finitely bounded rate (except for the special case when the HF reduces to the KF itself), while the estimation error of the KF does not possess such a guarantee.

In this work we propose a variant of the HF, called the time-local HF (TLHF), to avoid solving global constraints as in the HF. By analogy to the EnKF, we further introduce the ensemble TLHF (EnTLHF) for data assimilation in high dimensional systems. We show that the EnTLHF can be constructed based on the EnKF, and thus the computational complexity of the EnTLHF is in general comparable to that of the EnKF. We also show that some specific forms of the EnTLHF have connections with some EnKFs equipped with certain covariance inflation techniques. More generally, we show that an EnKF with a certain covariance inflation technique is in fact an EnTLHF.

The organization of this work is as follows. Section~\ref{sec:ps} presents the data assimilation problem and discusses its solutions in terms of the KF and the HF, respectively. Section \ref{sec:TLHF} introduces the TLHF as a variant of the HF, and its ensemble form, the EnTLHF. Section \ref{sec:specific_forms_of_EnTLHF} discusses some specific forms of the EnTLHF and shows their connections with some of the EnKF methods with covariance inflation. In section \ref{sec:examples}, we use some numerical examples to show the relative robustness of the TLHF/EnTLHF in comparison to the corresponding KF/EnKF method.

\section{Problem statement} \label{sec:ps}
We consider the state estimation problem in the following scenario:
\begin{subequations}  \label{ps} %
\begin{align}
 \label{ps_dyanmical_system} & \mathbf{x}_i  = \mathcal{M}_{i,i-1} \left( \mathbf{x}_{i-1} \right) + \mathbf{u}_{i}  \, ,  \\
  \label{ps_observation_system} &  \mathbf{y}_i  = \mathcal{H}_{i} \left( \mathbf{x}_{i} \right) + \mathbf{v}_{i} \, , \\
  \label{ps_dyanmical_noise} & \mathbb{E} \mathbf{u}_{i} =0;~\mathbb{E} (\mathbf{u}_{i}\mathbf{u}_{j}^T) = \delta_{ij} \mathbf{Q}_i \, ,\\
  \label{ps_observation_noise}& \mathbb{E} \mathbf{v}_{i} =0;~\mathbb{E} \left( \mathbf{v}_{i} \mathbf{v}_{j}^T \right) = \delta_{ij}\mathbf{R}_i \, ,\\
  \label{ps_uncorrelated_noise}& \mathbb{E} \left( \mathbf{u}_{i} \mathbf{v}_{j}^T \right) = 0 \quad \forall \, i, \, j \, .
\end{align}
\end{subequations}

Eqs.~(\ref{ps_dyanmical_system}) and (\ref{ps_observation_system}) represent the $m_x$-dimensional dynamical system and the $m_y$-dimensional observation system, respectively, where $\mathbf{x}_i$ denotes the $m_x$-dimensional state vector, $\mathbf{y}_i$ the corresponding $m_y$-dimensional observation, $\mathcal{M}_{i,i+1}: \mathbb{R}^{m_x} \rightarrow \mathbb{R}^{m_x}$ the transition operator, and $\mathcal{H}_{i}: \mathbb{R}^{m_x} \rightarrow \mathbb{R}^{m_y}$ the observation operator, at time $i$. For convenience, in this section we assume the systems in Eqs.~(\ref{ps_dyanmical_system}) and (\ref{ps_observation_system}) are linear so that $\mathcal{M}_{i,i+1}$ and $\mathcal{H}_{i}$ are two matrices. The estimation problem in the presence of nonlinearity will be addressed in terms of the EnTLHF, as will be discussed later.

Eqs.~(\ref{ps_dyanmical_noise})-(\ref{ps_uncorrelated_noise}) imply that the $m_x$-dimensional dynamical noise $\mathbf{u}_{i}$ and the $m_y$-dimensional observation noise $\mathbf{v}_{i}$ are uncorrelated white noise\footnote{The deduction will be similar in case that $\mathbf{u}_{i}$ and $\mathbf{v}_{i}$ are correlated colored noise. Readers are referred to, for example, \citet[ch. 7]{Simon2006} for the details.}, with zero mean and covariances $\mathbf{Q}_i$ and $\mathbf{R}_i$, respectively. $\delta_{ij}$ denotes the Kronecker delta function such that $\delta_{ij}=1$ if $i=j$, $\delta_{ij}=0$ otherwise. 

In what follows, we discuss two filtering approaches as the solutions of the above state estimation problem: The KF which is based on the criterion of minimizing the variance of the estimation error (equivalent to applying Bayes' rule to update background statistics as shown in \citealp[ch.~7]{Jazwinski1970}), and the optimal HF which is based on the criterion of minimizing the supremum (or maximum) of the ratio of the ``energy'' of the estimation error to the ``energy'' of the uncertainties in data assimilation (to be made clear shortly). In what follows we outline the main results of the KF and the HF. For more details, readers will be referred to appropriate references.

\subsection{Kalman filter as a solution}
The KF algorithm involves prediction and filtering steps, whose deductions can be found in, for example, \citet[ch.~3]{Simon2006}. When the KF is applied to assimilate the system in Eq.~(\ref{ps}), these steps are as follows.

\begin{itemize}

\item Prediction step: Propagate the analysis $\mathbf{x}_{i-1}^a$ at the $(i-1)$th cycle and the associated analysis error covariance $\mathbf{P}_{i-1}^a$ forward to obtain the background $\mathbf{x}_{i}^b$ at the next cycle (Eq.~(\ref{eq:KF_background})) and the associated background error covariance $\mathbf{P}_{i}^b$ (Eq.~(\ref{eq:KF_background_cov})), respectively.
\begin{subequations}\label{eq:KF_propagation}
\begin{align}
\label{eq:KF_background}& \mathbf{x}_{i}^b = \mathcal{M}_{i,i-1} \, \mathbf{x}_{i-1}^a \, , \\
\label{eq:KF_background_cov}& \mathbf{P}_{i}^b = \mathcal{M}_{i,i-1} \, \mathbf{P}_{i-1}^a \, \mathcal{M}_{i,i-1}^T + \mathbf{Q}_{i}\, .
\end{align}
\end{subequations}

\item Filtering step: With a new incoming observation $\mathbf{y}_i$, update the background $\mathbf{x}_{i}^b$ and the associated error covariance $\mathbf{P}_i^b$ to their  analysis counterparts, $\mathbf{x}_{i}^a$ (Eq.~(\ref{eq:KF_analysis})) and $\mathbf{P}_i^a$ (Eq.~(\ref{eq:KF_analysis_cov})), respectively.
\begin{subequations} \label{eq:KF_filtering}
\begin{align}
\label{eq:KF_analysis}& \mathbf{x}_i^a = \mathbf{x}_i^b + \mathbf{K}_i \left ( \mathbf{y}_i - \mathcal{H}_i \mathbf{x}_i^b \right ) \, , \\
\label{eq:KF_analysis_cov}& \mathbf{P}_i^a = \mathbf{P}_i^b -  \mathbf{K}_i \mathcal{H}_i \mathbf{P}_i^b \, , \\
\label{eq:KF_gain}& \mathbf{K}_i = \mathbf{P}_i^b \mathcal{H}_i^T ( \mathcal{H}_i \mathbf{P}_i^b \mathcal{H}_i^T + \mathbf{R}_i )^{-1} \, ,
\end{align}
\end{subequations}
\end{itemize}
where $\mathbf{K}_i$ in Eq.~(\ref{eq:KF_gain}) is the Kalman gain. Alternatively, Eqs.~(\ref{eq:KF_analysis_cov}) and (\ref{eq:KF_gain}) can be reformulated as \citep[ch.~6]{Simon2006}
\begin{subequations} \label{eq:alternative_KF_forms}
\begin{align}
\label{eq:alternative_analysis_analysis_cov}& (\mathbf{P}_i^a)^{-1} = (\mathbf{P}_i^b)^{-1} +   (\mathcal{H}_i)^T (\mathbf{R}_i )^{-1} \mathcal{H}_i \, , \\
\label{eq:alternative_KF_analysisKF_gain}& \mathbf{K}_i = \mathbf{P}_i^a \mathcal{H}_i^T (\mathbf{R}_i )^{-1} \, .
\end{align}
\end{subequations}
Eq.~(\ref{eq:alternative_KF_forms}) will be used to simplify the presentation of the HF later on.

To better understand the difference between the KF and the HF, we extend our discussion to a slightly more general situation. Instead of looking for an estimate of the system state $\mathbf{x}_{i}$, we are interested in estimating some $m_z$-dimensional linear transform $\mathbf{z}_i$ of $\mathbf{x}_i$, in terms of
\begin{equation} \label{eq: transform of x}
\mathbf{z}_i = \mathbf{L}_i \mathbf{x}_i \, ,
\end{equation}
with $\mathbf{L}_i$ being a user-defined $m_z \times m_x$ matrix. In Eq.~(\ref{eq: transform of x}), if $\mathbf{L}_i$ is the $m_x \times m_x$ identity matrix, then $\mathbf{z}_i$ reduces to $\mathbf{x}_i$ itself. But $\mathbf{L}_i$ can be something else in general. For instance, one may let $\mathbf{L}_i = [1, 0, \dotsb, 0] $ such that $\mathbf{z}_i$ is equivalent to the first element of $\mathbf{x}_i$.

In the KF, one can solve the estimation problem by minimizing the following cost function
\begin{equation} \label{eq: Kalman cost function}
J_{z}^{KF} (\mathbf{z}_0^a, \dotsb, \mathbf{z}_N^a) = \sum\limits_{i=0}^{N} J_{z,i}^{KF} = \sum\limits_{i=0}^{N} \mathbb{E} \Vert \mathbf{z}_i -\mathbf{z}_i^a \Vert_2^2
\end{equation}
with respect to the variables $\mathbf{z}_0^a, \dotsb, \mathbf{z}_N^a$. Here, $\mathbf{z}_i^a$ represents the analysis of the truth $\mathbf{z}_i$, while $J_{z,i}^{KF} \equiv \mathbb{E} \Vert \mathbf{z}_i -\mathbf{z}_i^a \Vert_2^2$, the cost function local to the $i$th cycle with respect to $\mathbf{z}_i$, is the variance of the estimation error $(\mathbf{z}_i -\mathbf{z}_i^a)$ at the $i$th assimilation cycle, with $\mathbb{E}$ being the expectation operator, and $\Vert \bullet \Vert_2$ denoting the $L_2$--norm. Given an $m_z$-dimensional vector $\mathbf{z}=(z_1,\dotsb,z_{m_z})^T$, $\Vert \mathbf{z} \Vert_2 = \sqrt{\mathbf{z}^T \mathbf{z}} =\sqrt{\sum\limits_{j=1}^{m_z} z_j^2}$.

The estimates of the KF are obtained by sequentially minimizing the local cost function $J_{z,i}^{KF}$ and then propagating the resulting analysis forward to yield the background at the next cycle. By treating $\mathbf{z}_i = \mathbf{L}_i \mathbf{x}_i$ as a system state, \emph{direct} sequential minimization of the local cost functions $J_{z,i}^{KF}$ is equivalent to running the KF recursively to assimilate the following modified systems
\begin{subequations}  \label{eq: modified systems} %
\begin{align}
 \label{eq: modified_dyanmical_system} & \mathbf{z}_i  = \left( \mathbf{L}_i \mathcal{M}_{i,i-1} \mathbf{L}_i^{-1} \right) \mathbf{z}_{i-1}+ \mathbf{L}_i \mathbf{u}_{i}  \, ,  \\
  \label{eq: modified_observation_system} &  \mathbf{y}_i  = \left( \mathcal{H}_{i} \mathbf{L}_i^{-1} \right) \mathbf{z}_{i} + \mathbf{v}_{i} \, ,
\end{align}
\end{subequations}
where $\mathbf{z}_i$ and $\mathbf{L}_i \mathbf{u}_{i}$ are the system state and dynamical noise at instant $i$, respectively, associated with the modified transition operator $\mathbf{L}_i \mathcal{M}_{i,i-1} \mathbf{L}_i^{-1}$, where $\mathbf{L}_i^{-1}$ may have to be treated as the pseudo-inverse of $\mathbf{L}_i$ in some circumstances. $\mathbf{y}_i$ and $\mathbf{v}_{i}$ still correspond to the observation vector and noise, respectively, but the observation operator becomes $\mathcal{H}_{i} \mathbf{L}_i^{-1}$.

Alternatively, one may solve the estimation problem in an \emph{indirect} way, but without changing the systems in assimilation. To this end, one specifies a new cost function
\begin{equation} \label{eq: cost function in x}
J_{x}^{KF} (\mathbf{x}_0^a, \dotsb, \mathbf{x}_N^a) = \sum\limits_{i=0}^{N} J_{x,i}^{KF} = \sum\limits_{i=0}^{N} \mathbb{E} \Vert \mathbf{x}_i -\mathbf{x}_i^a \Vert_2^2
\end{equation}
with respect to the variables $\mathbf{x}_0^a, \dotsb, \mathbf{x}_N^a$ to determine the optimal estimations by sequentially minimizing the local cost functions $J_{x,i}^{KF}$ (equivalent to running the KF recursively to assimilate the system in Eq.~(\ref{ps})), and then applies Eq.~(\ref{eq: transform of x}) to obtain the estimations $\mathbf{z}_0^a, \dotsb, \mathbf{z}_N^a$ of interest. 

The user-defined matrix $\mathbf{L}_i$ does not appear in Eq.~(\ref{eq: cost function in x}), which implies that in \emph{indirect} estimation, the estimates $\mathbf{x}_i^a$ of the KF do not depend on the choice of $\mathbf{L}_i$, so that by putting
\begin{equation} \label{eq:estimation of z}
\mathbf{z}_i^a = \mathbf{L}_i \mathbf{x}_i^a \, ,
\end{equation}
direct and indirect methods yield the same estimates $\mathbf{z}_i^a$. However, for the HF to be presented below, we will see that the choice of $\mathbf{L}_i$ does affect the estimates $\mathbf{x}_i^a$ in \emph{indirect} estimation, which is clearly different from the KF. 

\subsection{$\text{H}_{\infty}$ filter as a solution} \label{sec:HF_as_a_solution}

The HF (\citealp[ch.~11]{Simon2006} and the references therein) aims to provide robust, but not necessarily best, estimates. The main idea is to first recognize that in the system Eq.~(\ref{ps}), there are three possible sources that contribute to the uncertainties in data assimilation, namely, the uncertainties in the initial conditions, the model error and the observation error. Accordingly, during an assimilation time window $[0,N]$, these uncertainty sources are characterized by three uncertainty ``energy'' terms, defined as $\Vert \mathbf{x}_0 - \hat{\mathbf{x}}_0 \Vert_{\mathbf{\Delta}_0^{-1}}^2$, $ \sum\limits_{i=0}^{N} \Vert \mathbf{u}_i \Vert_{\mathbf{Q}_i^{-1}}^2$, and $\sum\limits_{i=0}^{N} \Vert \mathbf{v}_i\Vert_{\mathbf{R}_i^{-1}}^2$, respectively. For consistency, here we assume that an assimilation cycle starts from propagating an analysis forward through a dynamical system, so that the dynamical noise $\mathbf{u}_0$ is included in our discussion. Given a symmetric, positive semi-definite matrix $\mathbf{A}$, $\Vert \bullet \Vert_{\mathbf{A}}$ represents the weighted $L_2$--norm so that $\Vert \mathbf{z} \Vert_{\mathbf{A}} = \sqrt{\mathbf{z}^T \mathbf{A} \mathbf{z}}$ for a column vector $\mathbf{z}$.

Since the minimum variance criterion in the KF is consistent with Bayes's rule, it is customary to interpret the matrices $\mathbf{\Delta}_0$, $\mathbf{Q}_i$, and $\mathbf{R}_i$ as the covariances (or their estimates) of the corresponding random vectors in probabilistic language. In the context of the HF, however, the minimax rule is adopted instead, which does not have probabilistic interpretations for $\mathbf{\Delta}_0$, $\mathbf{Q}_i$, and $\mathbf{R}_i$ in general \citep{Simon2006}. Therefore, for conceptual distinction, in the context of the HF we call $\mathbf{\Delta}_0$, $\mathbf{Q}_i$, and $\mathbf{R}_i$ the \emph{uncertainty weight matrices} (uncertainty matrices for short) with respect to the initial conditions, the model error, and the observation error, respectively, and their inverses, $\mathbf{\Delta}_0^{-1}$, $\mathbf{Q}_i^{-1}$, and $\mathbf{R}_i^{-1}$, the corresponding \emph{information matrices}. The uncertainty matrices $\mathbf{\Delta}_0$, $\mathbf{Q}_i$, and $\mathbf{R}_i$ can be user-defined, although in practice it is convenient to let them be equal to their covariance counterparts (or the estimates) in the KF. Under this choice, the difference between the estimates of the KF and the HF becomes clear, as will be shown later.

To provide robust estimates, the HF requires that the total ``energy'' of the estimation errors, in terms of $\sum\limits_{i=0}^{N} \Vert \mathbf{z}_i -\mathbf{z}_i^a \Vert_{\mathbf{S}_i}^2$, be no larger than the sum of the above three uncertainty ``energy'' terms times a constant factor $1/\gamma$, i.e.,
\begin{equation} \label{eq:HF sub-optimal solution}
\sum\limits_{i=0}^{N} \Vert \mathbf{z}_i - \mathbf{z}_i^a \Vert_{\mathbf{S}_i}^2 \leq \dfrac{1}{\gamma}\left( \Vert \mathbf{x}_0 - \hat{\mathbf{x}}_0 \Vert_{\mathbf{\Delta}_0^{-1}}^2 + \sum\limits_{i=0}^{N} \Vert \mathbf{u}_i \Vert_{\mathbf{Q}_i^{-1}}^2 + \sum\limits_{i=0}^{N} \Vert \mathbf{v}_i\Vert_{\mathbf{R}_i^{-1}}^2 \right) \, .
\end{equation}
In (\ref{eq:HF sub-optimal solution}), the weight matrix $\mathbf{S}_i$ is also user-chosen, which can be used to guide the filter's behavior in estimation \citep[ch.~11]{Simon2006}. 

To solve (\ref{eq:HF sub-optimal solution}), we first define the following cost function $J_z^{HF}$
\begin{equation} \label{eq:HF cost function}
J_z^{HF} = \dfrac{\sum\limits_{i=0}^{N} \Vert \mathbf{z}_i -\mathbf{z}_i^a \Vert_{\mathbf{S}_i}^2}
{\Vert \mathbf{x}_0 - \hat{\mathbf{x}}_0 \Vert_{\mathbf{\Delta}_0^{-1}}^2 + \sum\limits_{i=0}^{N} \Vert \mathbf{u}_i \Vert_{\mathbf{Q}_i^{-1}}^2 + \sum\limits_{i=0}^{N} \Vert \mathbf{v}_i\Vert_{\mathbf{R}_i^{-1}}^2}
\end{equation}
which is a function of the variables $\mathbf{x}_0$, $\{ \mathbf{u}_i \}$, $\{ \mathbf{v}_i \}$ and $\{ \mathbf{z}_i^a \}$ \footnote{$\hat{\mathbf{x}}_0$ is the prior knowledge of $\mathbf{x}_0$, which is assumed to be available, and therefore not influenced by the filter designer.}. Then the inequality (\ref{eq:HF sub-optimal solution}) is equivalent to $J_z^{HF} \leq 1/\gamma$, except when the uncertainty ``energy'' term
\[
\Vert \mathbf{x}_0 - \hat{\mathbf{x}}_0 \Vert_{\mathbf{\Delta}_0^{-1}}^2 + \sum\limits_{i=0}^{N} \Vert \mathbf{u}_i \Vert_{\mathbf{Q}_i^{-1}}^2 + \sum\limits_{i=0}^{N} \Vert \mathbf{v}_i\Vert_{\mathbf{R}_i^{-1}}^2
\]
is zero. In this exceptional case, there still exists a HF solution to make the inequality (\ref{eq:HF sub-optimal solution}) hold. For example, if $\mathbf{\Delta}_0$, $\mathbf{Q}_i$ and $\mathbf{R}_i$ are all $+\infty$ so that their inverses are zero, we can choose $\mathbf{S}_i = \mathbf{0}$ so that the left hand side (lhs) of (\ref{eq:HF sub-optimal solution}) is also zero (for consistency, we define the products $\infty \cdot 0 = 0 \cdot \infty = 0$ to avoid possible problems). Similarly, if $\mathbf{x}_0 - \hat{\mathbf{x}}_0$, $\mathbf{u}_i$ and $\mathbf{v}_i$ are all zero, we have the ideal solution $\mathbf{z}_i^a = \mathbf{z}_i$ to make the lhs of (\ref{eq:HF sub-optimal solution}) become zero.

Let $\gamma^*$ be the value such that
\begin{equation} \label{eq:optimal HF solution}
\dfrac{1}{\gamma^*} = \inf\limits_{\{ \mathbf{z}_i^a \}}~\sup\limits_{\mathbf{x}_0, \{\mathbf{u}_i\},\{\mathbf{v}_i\}}~ J_z^{HF} \, ,~i \leq N ,
\end{equation}
where $\sup\limits_{\mathbf{x}_0, \{\mathbf{u}_i\},\{\mathbf{v}_i\}} J_z^{HF}$ calculates the supremum of the cost function $J_z^{HF}$ with respect to the variables $\mathbf{x}_0, \{\mathbf{u}_i\},\{\mathbf{v}_i \}$ (which is a function of $\{ \mathbf{z}_i^a \}$), and $\inf\limits_{\{ \mathbf{z}_i^a \}}$ evaluates the infimum with respect to $\{ \mathbf{z}_i^a \}$ along the previously obtained \emph{supremum plane} of $\mathbf{x}_0, \{\mathbf{u}_i\},\{\mathbf{v}_i \}$. We say that the optimal HF is achieved if $\gamma = \gamma^*$. In this sense, the evaluation of $\gamma^*$ is an application of the minimax rule, a strategy that aims to provide robust estimates and is different from its Bayesian counterpart \citep[chs.~4 and 5]{Burger1985}.

In practice, it is difficult to evaluate the exact value of $\gamma^*$, since, by the definition in Eq.~(\ref{eq:optimal HF solution}), $\gamma^*$ depends not only on the initial conditions and the dynamical and observation systems, but also on the length $N$ of the assimilation time window. A more convenient strategy is to choose a value $\gamma$ satisfying $1 / \gamma^* < 1/\gamma \leq +\infty$, so that it guarantees that there exists a (sub-optimal) HF solution $\{ \mathbf{z}_i^a \}$ satisfying the inequality (\ref{eq:HF sub-optimal solution}) \citep[ch.~11]{Simon2006}. In contrast, if $1/\gamma <1/\gamma^*$, then there is no guarantee (although it is not impossible) that there exists such a HF solution. For instance, in the extreme event with the variables $\mathbf{x}_0, \{\mathbf{u}_i\},\{\mathbf{v}_i \}$ located on their \emph{supremum plane}, then by the definition in Eq.~(\ref{eq:optimal HF solution}), there is no HF solution $\{ \mathbf{z}_i^a\}$ to make $J_z^{HF} \leq 1/\gamma < 1/\gamma^*$. For reference, hereafter we call $\gamma$ the \emph{ performance level} of the HF.

The inequality (\ref{eq:HF sub-optimal solution}) can be solved through dynamic constrained optimization, with Eqs.~(\ref{ps_dyanmical_system}) and (\ref{ps_observation_system}) being the constraints at different assimilation cycles. For details, readers are referred to \citet[ch.~11]{Simon2006}. For convenience of comparison, we also split the algorithm into prediction and filtering steps.

Prediction step: As in the KF, we also propagate the analysis forward to produce the background at the next cycle.
\begin{subequations}\label{eq:HF_propagation}
\begin{align}
\label{eq:HF_background}& \mathbf{x}_{i}^b = \mathcal{M}_{i,i-1} \, \mathbf{x}_{i-1}^a \, , \\
\label{eq:HF_background_cov}& \mathbf{\Delta}_{i}^b = \mathcal{M}_{i,i-1} \, \mathbf{\Delta}_{i-1}^a \, \mathcal{M}_{i,i-1}^T + \mathbf{Q}_{i}\, .
\end{align}
\end{subequations}

Filtering step: With a new observation $\mathbf{y}_i$, we update the background to the analysis.
\begin{subequations} \label{eq:HF_filtering}
\begin{align}
\label{eq:HF_analysis}& \mathbf{x}_i^a = \mathbf{x}_i^b + \mathbf{G}_i \left ( \mathbf{y}_i - \mathcal{H}_i \mathbf{x}_i^b \right ) \, , \\
\label{eq:HF_analysis_cov}& (\mathbf{\Delta}_i^a)^{-1} = (\mathbf{\Delta}_i^b)^{-1} +   (\mathcal{H}_i)^T (\mathbf{R}_i )^{-1} \mathcal{H}_i - \gamma \mathbf{L}_i^T \mathbf{S}_i \mathbf{L}_i \, , \\
\label{eq:HF_gain}& \mathbf{G}_i = \mathbf{\Delta}_i^a \mathcal{H}_i^T (\mathbf{R}_i )^{-1} \, ,
\end{align}
\end{subequations}
subject to the constraints
\begin{equation} \label{eq:HF_inequality_constraint}
(\mathbf{\Delta}_i^a)^{-1} = (\mathbf{\Delta}_i^b)^{-1} +   (\mathcal{H}_i)^T (\mathbf{R}_i )^{-1} \mathcal{H}_i - \gamma \mathbf{L}_i^T \mathbf{S}_i \mathbf{L}_i \geq 0 \, .
\end{equation}
Here $\mathbf{\Delta}_i$ denotes the uncertainty matrix, analogous to the covariance matrix $\mathbf{P}_i$ in the KF, and $\mathbf{G}_i$ represents the gain matrix, analogous to the Kalman gain $\mathbf{K}_i$. The symbol ``$\geq 0$'' means that $(\mathbf{\Delta}_i^a)^{-1}$ must be positive semi-definite. After obtaining $\mathbf{x}_i^a$, one can apply Eq.~(\ref{eq:estimation of z}) to estimate $\mathbf{z}_i$.

Comparing Eq.~(\ref{eq:HF_propagation}) with Eq.~(\ref{eq:KF_propagation}), one can see that the prediction steps of the KF and the HF are the same. Furthermore, the update formula Eq.~(\ref{eq:HF_analysis}) of the HF is a linear estimator as in the KF, but in general with a different gain matrix $\mathbf{G}_i$. However, if $\gamma =0$ in Eq.~(\ref{eq:HF_analysis_cov}), then Eqs.~(\ref{eq:HF_analysis}), (\ref{eq:HF_analysis_cov}) and (\ref{eq:HF_gain}) reduce to Eqs.~(\ref{eq:KF_analysis}), (\ref{eq:alternative_analysis_analysis_cov}) and (\ref{eq:alternative_KF_analysisKF_gain}), respectively. In this case, the HF is equivalent to the KF. Therefore, the KF can be considered as a special case of the HF with a performance level $\gamma=0$.

A further examination of Eq.~(\ref{eq:HF_filtering}) reveals that the user-defined matrix $\mathbf{L}_i$ is involved in the evaluation of $\mathbf{\Delta}_i^a$, hence those of $\mathbf{G}_i$ and $\mathbf{x}_i^a$. This implies that, unlike the KF, the choice of $\mathbf{L}_i$ affects the estimates of the HF. In addition, let
\begin{equation} \label{eq:KF_part_in_cov}
(\mathbf{\Sigma}_i^a)^{-1} = (\mathbf{\Delta}_i^b)^{-1} +   (\mathcal{H}_i)^T (\mathbf{R}_i )^{-1} \mathcal{H}_i \, ,
\end{equation}
then following Eq.~(\ref{eq:alternative_analysis_analysis_cov}), it is clear that $\mathbf{\Sigma}_i^a$ is an uncertainty matrix obtained by updating $\mathbf{\Delta}_i^b$ through the KF, while the information matrix
\begin{equation} \label{eg:connection_HFKF_cov}
(\mathbf{\Delta}_i^a)^{-1} = (\mathbf{\Sigma}_i^a)^{-1} - \gamma \mathbf{L}_i^T \mathbf{S}_i \mathbf{L}_i < (\mathbf{\Sigma}_i^a)^{-1} \, .
\end{equation}
Therefore, the HF appears more conservative, in the sense that it tends to make its analysis uncertainty matrix $\mathbf{\Delta}_i^a$ larger than its counterpart $\mathbf{\Sigma}_i^a$ in the KF, given the same background uncertainty matrix $\mathbf{\Delta}_i^b$. Consequently, when using Eq.~(\ref{eq:HF_analysis}) to update the background to the analysis, the HF allocates more weight to the observation $\mathbf{y}_i$ than the KF does, which may be preferred when the background is not very reliable because of the uncounted sources of uncertainties (similar arguments can also be found in, for example, \citealp{Jazwinski1970,VanLeeuwen2009}) \footnote{If, in contrast, the observation is very unreliable, then one may choose a negative value for $\gamma$ such that the background has relatively more weight in the update. In this work we confine ourselves to the scenario $\gamma \geq 0$.}. In fact, as will be shown later, this conservativeness exhibits connections to \begin{scriptsize}\end{scriptsize}covariance inflation techniques adopted in some EnKF methods.

The presence of the term $- \gamma \mathbf{L}_i^T \mathbf{S}_i \mathbf{L}_i$ in Eq.~(\ref{eq:HF_analysis_cov}) reflects the fundamental difference in the estimation strategies employed in the KF and the HF. As pointed out previously, the KF assumes that one has sufficient information of the statistical properties of both the dynamical and observation systems. Under this assumption, the KF looks for the best possible estimates under a certain optimality criterion (e.g., minimum variance or maximum likelihood). Therefore, the KF can achieve good performance if the statistical description of the assimilated system is sufficiently close to the reality. However, if there exists substantial deviation, then the KF may have poor performance, or even diverge (see \citealp{Schlee1967} for such examples). In contrast, the HF is more conservative than the KF since it only aims to provide robust, rather than best, estimates. The HF may perform worse than the KF if the statistical properties (e.g., pdf or moments) of the assimilated system are well characterized and close to the truth. However, if there exist more uncertainties, the HF can perform better than the KF (cf. \citealp[p.~358]{Simon2006} for some examples).

A further issue that may be of interest in practice is the choice of the term $- \gamma \mathbf{L}_i^T \mathbf{S}_i \mathbf{L}_i$, such that one may improve the performance of the filter under other measures, e.g., root mean squared errors (RMSE). This is possible, since the HF only accounts for the robustness property, and one has certain freedom in choosing the forms of the quantities $\gamma$ and $\mathbf{S}_i$ \footnote{Here we assume $\mathbf{L}_i$ is determined by the practical need given in Eq.~(\ref{eq: transform of x}).}, as long as $- \gamma \mathbf{L}_i^T \mathbf{S}_i \mathbf{L}_i$ satisfies the constraints in the HF. In other words, the robustness requirement in general does not yield unique choices of $\gamma$ and $\mathbf{S}_i$. One may impose other objective functions to the HF to improve its performance under the relevant measures. One such example is the mixed (nonlinear) Kalman and $H_{\infty}$ filter \citep[ch.~12]{Simon2006}, where one tries to minimize a least square cost function as in the KF, while imposing certain robustness on the estimates in the sense of Eq.~(\ref{eq:HF sub-optimal solution}). Other examples can be found in Anderson's works \citep{Anderson2007,Anderson2009} on adaptive covariance inflation in the EnKF, where the criterion that the pdf of the covariance inflation factor be maximized is used. This point will become clear after we establish the connection between robustness and covariance inflation in subsequent sections.

\section{Time-local ensemble $\text{H}_{\infty}$ filter} \label{sec:TLHF}

The HF has to satisfy the inequality constraints in (\ref{eq:HF_inequality_constraint}), which makes it challenging and inefficient for sequential data assimilation in certain circumstances. To see this, suppose that for $i=0,\dotsb, N$, the HF has an admissible solution $\{\mathbf{x}_i^a \}$, with a pre-specified performance level $\gamma$ satisfying all the inequality constraints in (\ref{eq:HF_inequality_constraint}). When extending the time horizon from $N$ to $N+1$, $\gamma$ may not satisfy the constraint at $N+1$. As a result, one has to choose a smaller value for $\gamma$ and re-start the assimilation in the new time window $[0, N+1]$, resulting in a different filter solution for $i=0,\dotsb, N$.

Alternatively, one may keep the solution between $i=0,\dotsb, N$ unchanged. From $N+1$, one uses a smaller value $\gamma'$ for estimation as long as it satisfies (\ref{eq:HF_inequality_constraint}). Once $\gamma'$ violates the constraint for a larger $N$, one adopts an even smaller performance level $\gamma''$, but still keeps the previously obtained estimates, and so on. In what follows, we extend this idea further. We propose a variant of the HF, called the time-local $\text{H}_{\infty}$ filter (TLHF), in which we impose a local cost function and adopt a local performance level $\gamma_i$ to solve a local constraint at each assimilation cycle.

\subsection{Time-local $\text{H}_{\infty}$ filter for linear systems}
We first introduce the TLHF for linear systems. The extension to nonlinear systems, analogous to the EnKF methods, will be presented in the next section.

In the TLHF, we define a local cost function
\begin{equation} \label{eq:local_HF_cost_function}
J_{z,i}^{HF} = \dfrac{\Vert \mathbf{z}_i -\mathbf{z}_i^a \Vert_{\mathbf{S}_i}^2}
{\Vert \mathbf{x}_i - \mathbf{x}_i^b \Vert_{(\mathbf{\Delta}_i^b)^{-1}}^2 + \Vert \mathbf{u}_i \Vert_{\mathbf{Q}_i^{-1}}^2 +  \Vert \mathbf{v}_i\Vert_{\mathbf{R}_i^{-1}}^2}
\end{equation}
with respect to the variables $\mathbf{x}_i$, $\mathbf{u}_i $, $\mathbf{v}_i $ and $\mathbf{z}_i^a$. Here $\mathbf{x}_i^b$ is the background and is fixed relative to the $i$th cycle ($\mathbf{x}_0^b = \hat{\mathbf{x}}_0$). For each $i$, $J_{z,i}^{HF}$ can be treated as a special case of the cost function $J_z^{HF}$ in Eq.~(\ref{eq:HF cost function}), as if $N=i=0$.

Analogous to (\ref{eq:HF sub-optimal solution}), we require
\begin{equation} \label{eq:local HF solution}
\Vert \mathbf{z}_i -\mathbf{z}_i^a \Vert_{\mathbf{S}_i}^2 \leq \dfrac{1}{\gamma_i}~\left( \Vert \mathbf{x}_i - \mathbf{x}_i^b \Vert_{(\mathbf{\Delta}_i^b)^{-1}}^2 + \Vert \mathbf{u}_i \Vert_{\mathbf{Q}_i^{-1}}^2 +  \Vert \mathbf{v}_i\Vert_{\mathbf{R}_i^{-1}}^2 \right) \, ,
\end{equation}
where $\gamma_i$ is a suitable local performance level, which satisfies
\begin{equation} \label{eq:implicit_constraint_of_local_performance_level}
\dfrac{1}{\gamma_i} > \dfrac{1}{\gamma_i^*} \equiv \inf\limits_{\mathbf{z}_i^a}~\sup\limits_{\mathbf{x}_i, \mathbf{u}_i,\mathbf{v}_i}~ J_{z,i}^{HF} \, ,
\end{equation}
with $1/\gamma_i^*$ being the minimax point of $J_{z,i}^{HF}$. In addition,  $\gamma_i$ also has to satisfy a local inequality constraint at time instant $i$ (to be shown later). Then for all such $\gamma_i$ ($0 \leq i \leq N$), we have
\begin{equation} \label{eq: sum of local HF solution}
\begin{split}
\sum\limits_{i=0}^N \Vert \mathbf{z}_i -\mathbf{z}_i^a \Vert_{\mathbf{S}_i}^2  & \leq \sum\limits_{i=0}^N \dfrac{1}{\gamma_i} \left( \Vert \mathbf{x}_i - \mathbf{x}_i^b \Vert_{(\mathbf{\Delta}_i^b)^{-1}}^2 + \Vert \mathbf{u}_i \Vert_{\mathbf{Q}_i^{-1}}^2 +  \Vert \mathbf{v}_i\Vert_{\mathbf{R}_i^{-1}}^2 \right) \\
& \leq \max\limits_i \{ \dfrac{1}{\gamma_i} \} \left( \sum\limits_{i=0}^N \Vert \mathbf{x}_i - \mathbf{x}_i^b \Vert_{(\mathbf{\Delta}_i^b)^{-1}}^2 + \sum\limits_{i=0}^N \Vert \mathbf{u}_i \Vert_{\mathbf{Q}_i^{-1}}^2 +  \sum\limits_{i=0}^N \Vert \mathbf{v}_i\Vert_{\mathbf{R}_i^{-1}}^2 \right) \, . \\
\end{split}
\end{equation}
This shows that the growth rate of the total ``energy'' of the estimation error is finite unless $\gamma_i =0$ for some $i$. Thus, the corresponding estimates $\{ \mathbf{x}_i^a \}$ may also provide a robust solution in the same sense as in the HF.

Eq.~(\ref{eq: sum of local HF solution}) bears a similar form to (\ref{eq:HF sub-optimal solution}), but also exhibits a clear difference. That is, in Eq.~(\ref{eq: sum of local HF solution}), the total ``energy'' of the uncertainties includes the contribution from the uncertainty in specifying the background at each assimilation cycle. In contrast, in (\ref{eq:HF sub-optimal solution}) the counterpart term only represents the contribution from the uncertainty in specifying the initial conditions. The extra terms in Eq.~(\ref{eq: sum of local HF solution}) provide a possibility to take into account the effect(s) of nonlinearity and/or other mechanisms that contribute to the estimation errors in the background, so that one does not have to significantly change the structure of the HF when extending it from linear systems to nonlinear ones. For example, in the presence of nonlinearity, there may exist extra uncertainties incurred by the chosen data assimilation algorithm itself (called algorithm uncertainty hereafter), including the linearization error when one uses the extended Kalman filter (EKF) to assimilate a nonlinear system, and more generally, the approximation error when one adopts an approximation scheme in assimilation, such as the sampling error in the EnKF, or the rank deficiency in a reduced rank filter. These potential uncertainties influence the estimations of the system states, but conceptually they might not belong to the uncertainties in specifying the dynamical or observation systems. Instead, one may treat them as the uncertainties in specifying the background, an extension of the uncertainties in specifying the initial conditions. With this treatment, one may apply the TLHF to a nonlinear system in the same way as it is applied to a linear system, while including the uncertainties due to the effect(s) of nonlinearity and/or any other error sources into the category of uncertainty in specifying the background\footnote{Like the extended KF, there also exists the extended HF containing more thorough treatment of nonlinearity (see, for example, \citealp{Shaked1995}), whose implementation, however, involves the derivative(s) of nonlinear functions and more sophisticated inequality constraints.}.

Following the same deductions in \citet[ch.~11]{Simon2006}, one can derive the steps of the TLHF as follows.

Prediction step:
\begin{subequations}\label{eq:local_HF_propagation}
\begin{align}
\label{eq:local_HF_background}& \mathbf{x}_{i}^b = \mathcal{M}_{i,i-1} \, \mathbf{x}_{i-1}^a \, , \\
\label{eq:local_HF_background_cov}& \mathbf{\Delta}_{i}^b = \mathcal{M}_{i,i-1} \, \mathbf{\Delta}_{i-1}^a \, \mathcal{M}_{i,i-1}^T + \mathbf{Q}_{i}\, .
\end{align}
\end{subequations}

Filtering step:
\begin{subequations} \label{eq:local_HF_filtering}
\begin{align}
\label{eq:local_HF_analysis}& \mathbf{x}_i^a = \mathbf{x}_i^b + \mathbf{G}_i \left ( \mathbf{y}_i - \mathcal{H}_i \mathbf{x}_i^b \right ) \, , \\
\label{eq:local_HF_analysis_cov}& (\mathbf{\Delta}_i^a)^{-1} = (\mathbf{\Delta}_i^b)^{-1} +   (\mathcal{H}_i)^T (\mathbf{R}_i )^{-1} \mathcal{H}_i - \gamma_i \mathbf{L}_i^T \mathbf{S}_i \mathbf{L}_i \, , \\
\label{eq:local_HF_gain}& \mathbf{G}_i = \mathbf{\Delta}_i^a \mathcal{H}_i^T (\mathbf{R}_i )^{-1} \, ,
\end{align}
\end{subequations}
subject to the constraint
\begin{equation} \label{eq:local_HF_inequality_constraint}
(\mathbf{\Delta}_i^a)^{-1} = (\mathbf{\Delta}_i^b)^{-1} +   (\mathcal{H}_i)^T (\mathbf{R}_i )^{-1} \mathcal{H}_i - \gamma_i \mathbf{L}_i^T \mathbf{S}_i \mathbf{L}_i \geq 0 \, .
\end{equation}

Thus compared with the HF, the TLHF only replaces the (global) performance level $\gamma$ with the local one $\gamma_i$ ($i=1,\dotsb,N$), without changing anything else.

\subsection{Ensemble time-local $\text{H}_{\infty}$ filter} \label{sec:EnTLHF_deduction}
The ensemble time-local $\text{H}_{\infty}$ filter (EnTLHF) is a straightforward analogue to the EnKF methods. Here the principal idea is that, at the prediction step, one uses the background ensemble, which is the propagation of the analysis ensemble from the previous cycle, to estimate the background and the associated uncertainty matrix. Then, one updates the background uncertainty matrix to the analysis one based on an EnKF method, calculates the corresponding Gain matrix of the EnTLHF, and computes the analysis mean and the associated uncertainty matrix (cf. Eq.~(\ref{eq:EnTLHF_filtering}) below).

Concretely, let $\mathbf{X}_i^b = \{ \mathbf{x}_{i,j}^b: \mathbf{x}_{i,j}^b = \mathcal{M}_{i,i-1} (\mathbf{x}_{i-1,j}^a), j=1,\dotsb,n \}$ be the $n$-member background ensemble at time instant $i$, which is the prediction of the analysis ensemble $\mathbf{X}_{i-1}^a = \{ \mathbf{x}_{i-1,j}^a, j=1,\dotsb,n \}$ at the previous cycle, with the transition operator $\mathcal{M}_{i,i-1}$ possibly being nonlinear.

At the prediction step of the EnTLHF:
\begin{subequations}\label{eq:EnTLHF_propagation}
\begin{align}
\label{eq:EnTLHF_background}& \hat{\mathbf{x}}_{i}^b = \text{mean} (\mathbf{X}_i^b) \, , \\
\label{eq:EnTLHF_background_cov}& \hat{\mathbf{\Delta}}_{i}^b = \text{cov} (\mathbf{X}_i^b) + \mathbf{Q}_i  \, ,
\end{align}
\end{subequations}
where $\hat{\mathbf{x}}_{i}^b$ and $\hat{\mathbf{\Delta}}_{i}^b$ denote the estimations of the background and the associated uncertainty matrix, respectively, and may be obtained through different ways in various EnKF methods. 

At the filtering step, one uses $\hat{\mathbf{\Delta}}_{i}^b$ in place of the background sample covariance in an EnKF method, and applies the corresponding formula in the EnKF method to update $\hat{\mathbf{\Delta}}_{i}^b$ to an uncertainty matrix $\hat{\mathbf{\Sigma}}_{i}^a$. Then by Eqs. (\ref{eq:KF_part_in_cov}) and (\ref{eg:connection_HFKF_cov}), one has
\begin{equation} \label{eq:connection_EnTLHFKF_cov}
\hat{\mathbf{\Delta}}_i^a = \left( \mathbf{I}_{m_x} - \gamma_i  \hat{\mathbf{\Sigma}}_{i}^a  (\mathbf{L}_i^T \mathbf{S}_i \mathbf{L}_i) \right)^{-1} \hat{\mathbf{\Sigma}}_{i}^a  \, ,
\end{equation}
where $\mathbf{I}_{m_x}$ denotes the $m_x$-dimensional identity matrix. Thus in general $\hat{\mathbf{\Delta}}_i^a$ can be evaluated based on the analysis update through an EnKF method. 

Substituting Eq.~(\ref{eq:connection_EnTLHFKF_cov}) into Eq.~(\ref{eq:local_HF_gain}), we have the estimated gain matrix
\begin{equation} \label{eq:connection_EnTLHFKF_gain}
\hat{\mathbf{G}}_i = \hat{\mathbf{\Delta}}_i^a \mathcal{H}_i^T (\mathbf{R}_i )^{-1} = \left( \mathbf{I}_{m_x} - \gamma_i  \hat{\mathbf{\Sigma}}_{i}^a  (\mathbf{L}_i^T \mathbf{S}_i \mathbf{L}_i) \right)^{-1} \hat{\mathbf{K}}_{i} \, ,
\end{equation}
where $\hat{\mathbf{K}}_{i} \equiv \hat{\mathbf{\Sigma}}_{i}^a \mathcal{H}_i^T (\mathbf{R}_i )^{-1}$ is the Kalman gain evaluated by an EnKF method, due to Eq.~(\ref{eq:alternative_KF_analysisKF_gain}). Note that for simplicity, in the above deduction we have assumed that the observation operator $\mathcal{H}_i$ is linear. In the case that  $\mathcal{H}_i$ is nonlinear, one could also derive an approximate formula of Eq.~(\ref{eq:connection_EnTLHFKF_gain}) (e.g., by linearizing $\mathcal{H}_i$ or using a Monte Carlo approach).

In particular, if one chooses $\gamma_i=0$ for $i=0,\dotsb,N$ in Eqs.~(\ref{eq:connection_EnTLHFKF_cov}) and (\ref{eq:connection_EnTLHFKF_gain}), then it is clear that $\hat{\mathbf{\Delta}}_i^a$ and $\hat{\mathbf{G}}_i$ are reduced to $\hat{\mathbf{\Sigma}}_{i}^a$ and $\hat{\mathbf{K}}_{i}$, their counterparts in the EnKF method, respectively. In this case, the EnTLHF is equivalent to the EnKF method without covariance inflation. If $\gamma_i>0$, the EnTLHF has connections to some EnKF methods with certain covariance inflation techniques, as will be discussed in \S\ref{sec:specific_forms_of_EnTLHF}.

For summary, at the filtering step of the EnTLHF:
\begin{subequations} \label{eq:EnTLHF_filtering}
\begin{align}
\label{eq:EnKF_output} & [\hat{\mathbf{\Sigma}}_{i}^a, \hat{\mathbf{K}}_{i}] = \text{EnKF} (\mathbf{X}_i^b, \mathbf{Q}_i, \mathcal{H}_i) \, , \\
\label{eq:EnTLHF_gain}& \hat{\mathbf{G}}_i = \left( \mathbf{I}_{m_x} - \gamma_i  \hat{\mathbf{\Sigma}}_{i}^a  (\mathbf{L}_i^T \mathbf{S}_i \mathbf{L}_i) \right)^{-1} \hat{\mathbf{K}}_{i} \, , \\
\label{eq:EnTLHF_analysis}& \hat{\mathbf{x}}_i^a = \hat{\mathbf{x}}_i^b + \hat{\mathbf{G}}_i \left ( \mathbf{y}_i - \mathcal{H}_i (\hat{\mathbf{x}}_i^b) \right ) \, , \\
\label{eq:EnTLHF_analysis_cov}& \hat{\mathbf{\Delta}}_i^a  = \left( \mathbf{I}_{m_x} - \gamma_i  \hat{\mathbf{\Sigma}}_{i}^a  (\mathbf{L}_i^T \mathbf{S}_i \mathbf{L}_i) \right)^{-1} \hat{\mathbf{\Sigma}}_{i}^a  \, ,
\end{align}
\end{subequations}
subject to the constraint
\begin{equation} \label{eq:EnTLHF_inequality_constraint}
(\hat{\mathbf{\Delta}}_i^a)^{-1} = (\hat{\mathbf{\Sigma}}_{i}^a)^{-1} - \gamma_i \mathbf{L}_i^T \mathbf{S}_i \mathbf{L}_i \geq 0 \, ,
\end{equation}
where Eq.~(\ref{eq:EnKF_output}) means that $\hat{\mathbf{\Sigma}}_{i}^a$ and $\hat{\mathbf{K}}_{i}$ are obtained through an EnKF method, given the information $\mathbf{X}_i^b$, $\mathbf{Q}_i$, and $\mathcal{H}_i$.

After obtaining $\hat{\mathbf{x}}_i^a$ and $\hat{\mathbf{\Delta}}_i^a$, one can draw an analysis ensemble of the system states, which preserves the estimate $\hat{\mathbf{x}}_i^a$ and the uncertainty matrix $\hat{\mathbf{\Delta}}_i^a$ \citep{Anderson-ensemble,Bishop-adaptive,Hoteit2002,Whitaker-ensemble}. This analysis ensemble is then integrated forward to start a new assimilation cycle.

\section{Some specific forms and their connections to covariance inflation} \label{sec:specific_forms_of_EnTLHF}

Here we show that some specific forms of the EnTLHF exhibit connections to some existing EnKF methods with covariance inflation. We again assume that the observation operator $\mathcal{H}_i$ is linear. This choice is made only for convenience in presenting the term $(\mathcal{H}_i)^T (\mathbf{R}_i )^{-1} \mathcal{H}_i$ below. The same results can also be obtained if $\mathcal{H}_i$ is nonlinear. 

From Eq.~({\ref{eq:EnTLHF_inequality_constraint}}), we have
\begin{equation} \label{eq:expanded_EnTLHF_inequality_constraint}
(\hat{\mathbf{\Delta}}_i^a)^{-1} = (\hat{\mathbf{\Delta}}_i^b)^{-1} +   (\mathcal{H}_i)^T (\mathbf{R}_i )^{-1} \mathcal{H}_i - \gamma_i \mathbf{L}_i^T \mathbf{S}_i \mathbf{L}_i
= (\hat{\mathbf{\Sigma}}_{i}^a)^{-1} - \gamma_i \mathbf{L}_i^T \mathbf{S}_i \mathbf{L}_i \geq 0 \, ,
\end{equation}
where the local performance level $\gamma_i$ and the information matrix $\mathbf{S}_i$ can be chosen by the designer. We consider some specific choices of the term $\gamma_i \mathbf{L}_i^T \mathbf{S}_i \mathbf{L}_i$, and derive the corresponding relations between $\hat{\mathbf{\Delta}}_i^a$ and $\hat{\mathbf{\Sigma}}_{i}^a$ (or $\hat{\mathbf{\Delta}}_i^b$). By treating $\hat{\mathbf{\Delta}}_i^a$ as the inflated covariance of $\hat{\mathbf{\Sigma}}_{i}^a$, the EnTLHF can be interpreted as an EnKF method equipped with a specific uncertainty inflation technique.

\subsection{Case $\gamma_i \mathbf{L}_i^T \mathbf{S}_i \mathbf{L}_i = c (\hat{\mathbf{\Delta}}_i^b)^{-1} $ for $0 \leq c \leq 1$} \label{sec:specific_form_seek}
This implies that
\begin{equation}
\mathbf{S}_i = \dfrac{c}{\gamma_i} (\mathbf{L}_i \mathbf{L}_i^T)^{-1} \mathbf{L}_i (\hat{\mathbf{\Delta}}_i^b)^{-1} (\mathbf{L}_i^T) (\mathbf{L}_i \mathbf{L}_i^T)^{-1} \, ,
\end{equation}
and
\begin{equation}
(\hat{\mathbf{\Delta}}_i^a)^{-1} = (1 - c) (\hat{\mathbf{\Delta}}_i^b)^{-1} +   (\mathcal{H}_i)^T (\mathbf{R}_i )^{-1} \mathcal{H}_i \, .
\end{equation}
In particular, if $\mathbf{L}_i$ is an identity matrix, $\mathbf{S}_i \propto (\hat{\mathbf{\Delta}}_i^b)^{-1}$, so that the uncertainty matrix $\mathbf{S}_i^{-1}$ is proportional to the background uncertainty matrix $\hat{\mathbf{\Delta}}_i^b$. For convenience, we label this specific form by I-BG, and call $c$ the \emph{performance level coefficient} (PLC).

If one ignores $\mathbf{Q}_i$ in Eq.~(\ref{eq:EnTLHF_background_cov}) (or if there is no dynamical noise term so that $\mathbf{Q}_i = \mathbf{0}$), and decomposes $\hat{\mathbf{\Delta}}_i^b$ and $\hat{\mathbf{\Delta}}_i^a$ as $\hat{\mathbf{\Delta}}_i^b = \mathbf{\Gamma}_i \mathbf{U}_i^b \mathbf{\Gamma}_i^T$ and $\hat{\mathbf{\Delta}}_i^a = \mathbf{\Gamma}_i \mathbf{U}_i^a \mathbf{\Gamma}_i^T$, respectively, where $\mathbf{\Gamma}_i$, $\mathbf{U}_i^b$ and $\mathbf{U}_i^a$ are suitable matrices, then it can be shown that
\begin{equation} \label{eq:seek_update}
(\mathbf{U}_i^a)^{-1} = (1 - c) (\mathbf{U}_i^b)^{-1} +   (\mathcal{H}_i \mathbf{\Gamma}_i)^T (\mathbf{R}_i )^{-1} (\mathcal{H}_i \mathbf{\Gamma}_i) \, ,
\end{equation}
which is equivalent to the singular evolutive extended Kalman (SEEK) filter \citep{Hoteit2002,Pham1998} with a forgetting factor $(1 - c)$.

\subsection{Case $\gamma_i \mathbf{L}_i^T \mathbf{S}_i \mathbf{L}_i = c~ [ (\hat{\mathbf{\Delta}}_i^b)^{-1} + (\mathcal{H}_i)^T (\mathbf{R}_i )^{-1} \mathcal{H}_i ] = c~ (\hat{\mathbf{\Sigma}}_i^a)^{-1}$ for $0 < c \leq 1$} \label{sec:specific_form_EnKF}
This implies that
\begin{equation}
\mathbf{S}_i = \dfrac{c}{\gamma_i} (\mathbf{L}_i \mathbf{L}_i^T)^{-1} \mathbf{L}_i ( \hat{\mathbf{\Sigma}}_i^a)^{-1} (\mathbf{L}_i^T) (\mathbf{L}_i \mathbf{L}_i^T)^{-1} \, ,
\end{equation}
and
\begin{equation} \label{eq:conventional_covariance_inflation}
(\hat{\mathbf{\Delta}}_i^a)^{-1} = (1 - c) (\hat{\mathbf{\Sigma}}_i^a)^{-1} \, .
\end{equation}
In particular, if $\mathbf{L}_i$ is an identity matrix, $\mathbf{S}_i \propto (\hat{\mathbf{\Sigma}}_i^a)^{-1}$, so that the uncertainty matrix $\mathbf{S}_i^{-1}$ is proportional to the analysis uncertainty matrix $\hat{\mathbf{\Sigma}}_i^a$ that is updated from $\hat{\mathbf{\Delta}}_i^b$ through an EnKF method. We label this specific form by I-ANA.

Eq.~(\ref{eq:conventional_covariance_inflation}) means that
\begin{equation}
\hat{\mathbf{\Delta}}_i^a= (1 - c)^{-1} \hat{\mathbf{\Sigma}}_i^a \, ,
\end{equation}
which is equivalent to the covariance inflation technique used in \cite{Anderson-Monte,Whitaker-ensemble} and other similar works if one lets $1/(1 - c) = (1+\delta)^2$, with $\delta$ being the inflation factor. Note that, in the EnKF method, one may first update the background mean to the analysis counterpart, and then generate an analysis ensemble, whose error covariance is (implicitly) inflated. In doing so, covariance inflation does not affect the computation of the Kalman gain at the same cycle. Instead, it affects the Kalman gain in the next cycle, since the background covariance at the next cycle will be inflated by conducting covariance inflation. In contrast, in the EnTLHF, the Gain matrix is directly affected by the analysis uncertainty matrix at the same assimilation cycle.

\subsection{Case $\mathbf{L}_i^T \mathbf{S}_i \mathbf{L}_i = \mathbf{I}_{m_x}$} \label{sec:specific_form_local_EnKF}
This implies that
\begin{equation}
\mathbf{S}_i =(\mathbf{L}_i \mathbf{L}_i^T)^{-1} \, ,
\end{equation}
and
\begin{equation} \label{eq:eigenvalue_covariance_inflation}
(\hat{\mathbf{\Delta}}_i^a)^{-1} = (\hat{\mathbf{\Sigma}}_i^a)^{-1} - \gamma_i \mathbf{I}_{m_x} \, .
\end{equation}
In particular, if $\mathbf{L}_i$ is an identity matrix, $\mathbf{S}_i = \mathbf{I}_{m_x}$, so that the uncertainty matrix $\mathbf{S}_i^{-1}$ is also equal to the identity matrix $\mathbf{I}_{m_x}$. We label this specific form by I-MTX.

Suppose that the analysis covariance $\hat{\mathbf{\Sigma}}_i^a$ obtained by an EnKF method is decomposed as
\begin{equation} \label{eq:EnKF_analysis_cov_svd}
\hat{\mathbf{\Sigma}}_i^a = \mathbf{E}_i^a \mathbf{D}_i^a (\mathbf{E}_i^a)^T
\end{equation}
through a singular value decomposition (SVD), where $\mathbf{D}_i^a = \text{diag} (\sigma_{i,1}, \dotsb, \sigma_{i,m_x})$ is a diagonal matrix consisting of the eigenvalues $\sigma_{i,j}$ of $\hat{\mathbf{\Sigma}}_i^a$ ($j=1,\dotsb,m_x$), which are arranged in a non-ascending order, i.e., $\sigma_{i,j} \ge \sigma_{i,k} \ge 0$ for $j<k$, and $\mathbf{E}_i^a = \left[\mathbf{e}_{i,1}, \dotsb,  \mathbf{e}_{i,m_x} \right]$ is the matrix consisting of the corresponding (orthonormal) eigenvectors $\mathbf{e}_{i,j}$ ($j=1,\dotsb,m_x$). Then in order to make $(\hat{\mathbf{\Delta}}_i^a)^{-1}$ in Eq.~(\ref{eq:eigenvalue_covariance_inflation}) positive definite, it is sufficient to let
\begin{equation}
\sigma_{i,1}^{-1} - \gamma_i > 0 \, ,
\end{equation}
which means
\begin{equation} \label{eq:specific_constraint_eigenvalue}
\gamma_i < \dfrac{1}{\sigma_{i,1}} \, ,
\end{equation}
i.e., $\gamma_i$ is less than the inverse of the maximum eigenvalue of $\hat{\mathbf{\Sigma}}_i^a$. We can also introduce a PLC to the filter in this case by writing $\gamma_i = c/\sigma_{i,1}$, with $0 \leq c <1$.

Given a suitable value of $\gamma_i$, $\hat{\mathbf{\Delta}}_i^a$ can be evaluated as follows. Using Eq.~(\ref{eq:eigenvalue_covariance_inflation}), one has
\begin{equation}
\begin{split}
(\hat{\mathbf{\Delta}}_i^a)^{-1} & = (\hat{\mathbf{\Sigma}}_i^a)^{-1} - \gamma_i \mathbf{I}_{m_x} \\
& = \mathbf{E}_i^a (\mathbf{D}_i^a)^{-1} (\mathbf{E}_i^a)^T - \gamma_i \mathbf{E}_i^a (\mathbf{E}_i^a)^T \\
& = \mathbf{E}_i^a (\mathbf{\Lambda}_i^a)^{-1} (\mathbf{E}_i^a)^T \, ,
\end{split}
\end{equation}
where $\mathbf{\Lambda}_i^a$ is a diagonal matrix consisting of the eigenvalues $\eta_{i,j}$ ($j=1,\dotsb,m_x$), with
\begin{equation} \label{eq:inflated_eigenvalue}
\eta_{i,j} = \dfrac{\sigma_{i,j}}{1-\gamma_i \sigma_{i,j}} = \dfrac{\sigma_{i,j}}{1-c~ \sigma_{i,j} / \sigma_{i,1}} \, .
\end{equation}
Thus one has
\begin{equation} \label{eq:inflated_covariance_through_eigenvalue}
\hat{\mathbf{\Delta}}_i^a = \mathbf{E}_i^a \mathbf{\Lambda}_i^a (\mathbf{E}_i^a)^T \, .
\end{equation}

Through Eqs.~(\ref{eq:inflated_eigenvalue}) and (\ref{eq:inflated_covariance_through_eigenvalue}), one can see that, the analysis uncertainty matrix $\hat{\mathbf{\Delta}}_i^a$ obtained by the EnTLHF has a similar structure to that of $\hat{\mathbf{\Sigma}}_i^a$ (cf. Eq.~(\ref{eq:EnKF_analysis_cov_svd})). The eigenvalues of $\hat{\mathbf{\Delta}}_i^a$ are the inflations of the corresponding eigenvalues $\sigma_{i,j}$ of $\hat{\mathbf{\Sigma}}_i^a$ if $\sigma_{i,j} >0$, or remain unchanged if $\sigma_{i,j} =0$.

\noindent{\textbf{Remarks:}}
A similar eigenvalue inflation technique was used in \cite{Ott-local}, where the authors increased all the eigenvalues of $\hat{\mathbf{\Sigma}}_i^a$ by a constant $\epsilon_i$ (relative to the $i$th cycle), so that after inflation, the eigenvalues of $\hat{\mathbf{\Delta}}_i^a$ are equal to $\sigma_{i,j} + \epsilon_i$, but with the same associated eigenvectors as in $\hat{\mathbf{\Sigma}}_i^a$. One may establish the connection of this inflation technique to the EnTLHF by setting $\gamma_i \mathbf{L}_i^T \mathbf{S}_i \mathbf{L}_i = (\hat{\mathbf{\Sigma}}_{i}^a)^{-1} - (\hat{\mathbf{\Sigma}}_{i}^a+ \epsilon_i \mathbf{I}_{m_x})^{-1}$ in Eq.~(\ref{eq:expanded_EnTLHF_inequality_constraint}).  Nominally, in light of Eq.~(\ref{eq:expanded_EnTLHF_inequality_constraint}), one may establish the connection of any inflation technique to the EnTLHF by solving the equation
\[
\gamma_i \mathbf{L}_i^T \mathbf{S}_i \mathbf{L}_i = (\hat{\mathbf{\Sigma}}_{i}^a)^{-1} -  (\hat{\mathbf{\Delta}}_i^a)^{-1}.
\]

As one more example, one may consider the scenario $\gamma_i \mathbf{L}_i^T \mathbf{S}_i \mathbf{L}_i = (\hat{\mathbf{\Sigma}}_{i}^a)^{-1} -  (\hat{\mathbf{\Delta}}_i^a)^{-1} = c (\mathcal{H}_i)^T (\mathbf{R}_i )^{-1} \mathcal{H}_i $ for $0 < c \leq 1$. This leads to
\begin{equation}
\mathbf{S}_i = \dfrac{c}{\gamma_i} (\mathbf{L}_i \mathbf{L}_i^T)^{-1} \mathbf{L}_i (\mathcal{H}_i)^T (\mathbf{R}_i )^{-1} \mathcal{H}_i (\mathbf{L}_i^T) (\mathbf{L}_i \mathbf{L}_i^T)^{-1} \, ,
\end{equation}
and
\begin{equation} \label{eq:observation_cov_inflation}
(\hat{\mathbf{\Delta}}_i^a)^{-1} = (\hat{\mathbf{\Delta}}_i^b)^{-1} + (1 - c) (\mathcal{H}_i)^T (\mathbf{R}_i )^{-1} \mathcal{H}_i \, .
\end{equation}
To our knowledge, covariance inflation based on Eq.~(\ref{eq:observation_cov_inflation}) is not used in the literature, possibly because it is more natural to conduct covariance inflation through the background or analysis ensemble of the system state, rather than through the ensemble of observation. However, Eq.~(\ref{eq:observation_cov_inflation}) might provide an alternative point of view to explain the under-performance of the stochastic EnKF \citep{Burgers-analysis,Evensen-sequential,Evensen-assimilation,Houtekamer1998} in comparison to the deterministic ones \citep{Anderson-ensemble,Bishop-adaptive,Whitaker-ensemble} in certain situations (e.g., \citealp{Whitaker-ensemble}). In the stochastic EnKF, one generates an ensemble of surrogate observations based on the observation distribution. Due to the effect of small sample size, the sample covariance of the surrogate observations will under-represent the original covariance of observation. This is equivalent to letting $1-c>1$ in Eq.~(\ref{eq:observation_cov_inflation}) (i.e., $c<0$), which implies the choice of $\gamma_i < 0$ in Eq.~(\ref{eq:expanded_EnTLHF_inequality_constraint}). As discussed previously, the negativeness of $\gamma_i$ means that, instead of being conservative, the filter designer is confident in the estimation accuracy of the background. Hence when updating the background to the analysis, more weight is allocated to the background, rather than to the observation, which may deteriorate the performance of the filter if there exist more uncertainties in the background than in the observation.

\section{Numerical examples} \label{sec:examples}

We conduct a series of numerical experiments to assess the relative robustness of the TLHF/EnTLHF in comparison to the corresponding KF/EnKF method without inflation. In all experiments, we estimate the full state vectors so that the transform matrix $\mathbf{L}_i=\mathbf{I}_{m_x}$.

\subsection{Experiments with a linear model}
In the first set of experiments, we consider a one-dimensional regression model, governed by the equation
\begin{equation} \label{eq:true_one_dimensional_regression_model}
x_{i+1} = 1+ 0.5 x_i - 0.1 x_i^2 + f(x_i;k,h,d) + u_i \, ,
\end{equation}
where the random variable $u_i$ follows the Gaussian distribution with mean zero and variance $1$ (denoted by $u_i \sim \text{N} (u_i:0,1)$), $f(x_i;k,h,d)$ is a discrete carbox function, starting at the time index $k$, with a jump height $h$ and a width $d$, i.e.,
\begin{equation} \label{eq:discrete_carbox}
f(x_i;k,h,d) = \begin{cases}
h,~\text{for}~ i = k,k+1,\dotsb, k+d-1 \, , \\
0, \text{otherwise} \, ,
\end{cases}
\end{equation}
In the experiment we let four jumps occur with $k=200$, $400$, $600$ and $800$, respectively, with the objective to verify that the TLHF can perform (almost) equally well for the jumps occurred at different times. For each jump, we fix $d=20$, but let the jump height $h$ be $10$ or $30$. For illustration, a time series generated by Eq.~(\ref{eq:true_one_dimensional_regression_model}) with $h=10$ is shown in Fig.~(\ref{fig:simAR1_time_series}).

We suppose that the associated observation system is characterized by
\begin{equation} \label{eq:one-dimensional_observation_system}
y_i = x_i + v_i \, ,
\end{equation}
where $v_i \sim \text{N}(v_i:0,1)$. In the experiment, we let the time index $i=1,2,\dotsb,1000$, and record the observation every time step. To reduce statistical fluctuations in estimation, we conduct the experiment for $100$ times, each time with different initial conditions (drawn at random), hence different truths and observations.

For assimilation, we suppose that one uses the following imperfect model
\begin{equation} \label{eq:imperfect_one_dimensional_regression_model}
x_{i+1} = 1+ 0.5 x_i + u_i \, ,
\end{equation}
as the dynamical system, which ignores the nonlinear term $-0.1 x_i^2$ and fails to capture the abrupt change of regime. Because of the linearity in Eqs.~(\ref{eq:one-dimensional_observation_system}) and (\ref{eq:imperfect_one_dimensional_regression_model}), we can apply both the KF and the TLHF to assimilate the system. Given the covariance $\Sigma_i$ (of a scalar variable) evaluated by the KF at the $i$th cycle, in the TLHF one only needs to replace it by $\Delta_i = \Sigma_i/(1-\gamma_i \Sigma_i) = \Sigma_i/(1-c)$ (with $\gamma_i = c / \Sigma_i, c \in [0, 1)$ in the scalar system) and change the gain matrix accordingly, while the other steps are the same as those in the KF. Clearly, in the scalar system, the specific form I-ANA in \S\ref{sec:specific_forms_of_EnTLHF}\ref{sec:specific_form_EnKF} is equivalent to the specific form I-MTX in \S\ref{sec:specific_forms_of_EnTLHF}\ref{sec:specific_form_local_EnKF}.

We use the average root mean squared error (average RMSE) to measure the performance of the filter. For an $m_x$-dimensional system, the RMSE $e_i$ of an estimate $\hat{\mathbf{x}}_i = (\hat{x}_{i,1}, \dotsb, \hat{x}_{i,m_x})^T$ with respect to the state vector $\mathbf{x}_i = (x_{i,1}, \dotsb, x_{i,m_x})^T$ at time instant $i$ is defined as
\begin{equation} \label{eq:def_of_rmse}
e_i = \dfrac{\Vert \hat{\mathbf{x}}_i - \mathbf{x}_i  \Vert_2}{\sqrt{m_x}} = \sqrt{\dfrac{1}{m_x} \sum\limits_{j=1}^{m_x} (\hat{x}_{i,j} - x_{i,j})^2} ~ .
\end{equation}
The average RMSE (RMSE for short) $\hat{e}_i$ at time instant $i$ over $100$ simulations is defined as $\hat{e}_i =\sum_{j=1}^{100} e_i^j/100$, where $e_i^j$ denotes the RMSE at time instant $i$ in the $j$th simulation. We also define the time mean RMSE $\hat{e}$ as the average of the (average) RMSE $\hat{e}_i$ over the time horizon (assimilation time window) $[1,\dotsb,N]$, i.e., $\hat{e} = \sum_{i=1}^{N} \hat{e}_i/N$ ($N = 1000$ in this experiment).

Fig.~\ref{fig:simAR1_KF_varying_h1030} plots the RMSE of the KF over the time horizon $[1,1000]$ in the case $h=10$ (upper panel), and that in the case $h=30$ (lower panel). In both cases, the KF achieves a relatively low RMSE during the period without any abrupt jump. However, when the abrupt jumps occur, the RMSE of the KF rises sharply in response.

Figs.~\ref{fig:simAR1_HF_KF_varying_h1030_bg_c01} -- \ref{fig:simAR1_HF_KF_varying_h1030_bg_c09} plot the RMSE differences between the TLHF of I-BG with different PLC values and the KF. Throughout this work, we use the RMSEs of the KF as the baselines, and the RMSE differences are defined as the RMSEs of the TLHF subtracted by the corresponding ones of the KF. In all these figures, the upper panels correspond to the case $h=10$, and the lower ones the case $h=30$. At $c=0.1$ (Fig.~\ref{fig:simAR1_HF_KF_varying_h1030_bg_c01}), when there is no abrupt jump, the RMSEs of the TLHF and the KF are nearly indistinguishable, so that their RMSE differences are almost zero. However, when the abrupt jumps appear, the RMSEs of the TLHF do not rise as sharply as those of the KF, so that their RMSE differences become negative, suggesting that the TLHF has relatively more robust performance than the KF during the abrupt jumps. At $c=0.5$ (Fig.~\ref{fig:simAR1_HF_KF_varying_h1030_bg_c05}), the RMSE differences during the periods with the abrupt jumps become larger, while those during the periods without any abrupt jump remain close to zero. Further increasing $c$ to $0.9$ (Fig.~\ref{fig:simAR1_HF_KF_varying_h1030_bg_c09}), the performance of the TLHF becomes remarkably better than the KF during the periods with abrupt jumps. The RMSEs of the TLHF appear insensitive to the presence of the abrupt jumps, which is not the case for the KF. However, there is also a price for the TLHF to achieve this. During the periods without the abrupt jumps, the TLHF performs worse than the KF so that their RMSE differences are slightly above zero. Moreover, the divergence of the TLHF is spotted for time indices $i > 870$. The occurrence of the divergence is possibly due to the fact that the PLC is too large, so that $1/\gamma_i$ becomes less than the minimum threshold $1/\gamma^{*}$ defined in Eq.~(\ref{eq:optimal HF solution}). As discussed in \S\ref{sec:ps}\ref{sec:HF_as_a_solution}, in such situations there is no guarantee to attain a TLHF solution that satisfies the inequality in (\ref{eq:local HF solution}). Instead, divergence of the filter solution may occur as observed in the experiment.

Figs.~\ref{fig:simAR1_HF_KF_varying_h1030_ana_c01} -- \ref{fig:simAR1_HF_KF_varying_h1030_ana_c06} show the RMSE differences between the TLHFs of I-ANA and I-MTX  with three different PLC values (equivalent to each other in scalar systems), and the KF. Similar results are observed. At $c=0.1$, the RMSEs of the TLHF and the KF are almost indistinguishable when there is no abrupt jump, so that their RMSE differences are very close to zero. The TLHF performs better again than the KF when the abrupt jumps occur. At a larger PLC value, $c=0.4$, the TLHF performs remarkably better than the KF during the period of abrupt jumps, but at the cost of slightly worse performance than the KF during the period without any abrupt jump. When further increasing $c$ to $0.6$, the performance of the TLHF deteriorates in comparison with the choice $c=0.4$. More investigations (not reported here) show that a larger value ($c>0.6$) leads to even worse performance.

To summarize, our experiment results show that, for a relatively small PLC, the KF and the TLHF have close performance. This is expected, since the TLHF with $c=0$ reduces to the KF as we have noted in \S\ref{sec:TLHF}. As $c$ increases, the TLHF exhibits a better performance than the KF when there are relatively large uncertainties. However, when there only exist relatively small uncertainties in assimilation, a too large $c$ (hence too much uncertainty inflation) may also make the TLHF appear over-conservative and deteriorate the filter performance (or even diverge). This is because, with relatively small uncertainties, the backgrounds also provide useful information, and thus should not be under-weighted. To mitigate this problem, one possible strategy is to use a relatively small value of $c$ to make the TLHF less conservative when there only exists relatively small uncertainties, and a larger one when there exhibit more uncertainties. This is essentially a strategy that conducts adaptive covariance inflation, as has already been investigated in some works, for example, \citet{Anderson2007,Anderson2009,Hoteit2002,Hoteit2004-adaptively}. From our earlier discussion in \S~\ref{sec:TLHF}\ref{sec:HF_as_a_solution}, the adaptive inflation problem can be solved under the framework of the HF with an additional optimality criterion (e.g, minimum variance or maximum likelihood), which will be investigated in the future.

\subsection{Experiments with a nonlinear model}
In the second set of experiments, we consider the $40$-dimensional Lorenz and Emanuel model \citep{Lorenz-optimal} (LE98 model hereafter), whose governing equations are given by
\begin{equation} \label{LE98}
\frac{dx_i}{dt} = \left( x_{i+1} - x_{i-2} \right) x_{i-1} - x_i + F(t), \, i=1, \dotsb, 40.
\end{equation}
The quadratic terms simulate advection, the linear term represents internal dissipation, and $F$ acts as the external forcing term \citep{Lorenz-predictability}. For consistency, we define $x_{-1}=x_{39}$, $x_{0}=x_{40}$, and $x_{41}=x_{1}$. We suppose that the true value of the parameter $F$ is $8$ for $t \geq 0$, but in assimilation one may choose other values for $F$, which thus yields a potential parameter mismatch. In our experiments we consider two scenarios, with $F=6$ and $F=8$, respectively.

We use the fourth-order Runge-Kutta method to integrate (and discretize) the system from time $0$ to $250$, with a constant integration step of $0.05$ (overall $5001$ integration steps). We then use the following system
\begin{equation}
\mathbf{y}_i = \mathbf{x}_i + \mathbf{v}_i
\end{equation}
to record the observation of the state vector $\mathbf{x}_i \equiv (x_{i,1},x_{i,2},\dotsb,x_{i,40})^T$ at time instant $i$, where $\mathbf{v}_i$ follows the Gaussian distribution $N(\mathbf{v}_i:\mathbf{0},\mathbf{I}_{40})$, with $\mathbf{I}_{40}$ being the $40$-dimensional identity matrix. The observations are made for every $4$ integration steps.

We use the ensemble transform Kalman filter (ETKF) \citep{Bishop-adaptive} to construct the EnTLHF. The ETKFs with I-BG and I-ANA are constructed by inflating the background ensembles and the analysis ensembles, respectively, in a similar way to that in \cite{Anderson-Monte,Whitaker-ensemble}. To construct the ETKF with I-MTX, one needs to evaluate the analysis covariances, conduct SVDs, and then inflate the associated eigenvalues. In high-dimensional systems, conducting SVDs on the analysis covariances makes the ETKF with I-MTX computationally less efficient than its I-BG and I-ANA counterparts. However, it is possible to implement the I-MTX form in the SEEK filter \citep{Hoteit2002,Pham1998} without significant increase of computational cost, since in this case all such SVDs can be conducted on the matrices updated by Eq.~(\ref{eq:seek_update}), whose dimension is determined by the ensemble size in assimilation.

In our experiments we let the ensemble size $n=10$ and vary the PLC values. To reduce statistical fluctuations, for each PLC value $c$ we repeat the experiments for $20$ times, each time with a randomly drawn initial background ensemble (with $10$ members). In practice, it is customary to introduce covariance localization to the ETKF in order to improve the filter performance \citep{Hamill2009,VanLeeuwen2009}. Since in our experiments our objective is to assess the relative robustness of the EnTLHF, we choose not to conduct covariance localization to avoid complicating the analysis of our experiment results. In what follows, we examine the time mean RMSE of the EnTLHF as a function of the PLC value $c$, with $c \in [0,0.1,0,2,\dotsb,0.9]$. The ETKF is treated as a special case of the EnTLHF with $c = 0$.

Fig.~\ref{fig:simL96_EnHF_nLoop20_bg_timeAvgRmse} plots the time mean RMSEs of the ETKF with I-BG. The result in case of $F=6$ is marked with the dash-dotted line, and that in the case of $F=8$ with the dotted one. When $F=6$, the time mean RMSE appears to be a monotonically decreasing function with respect to $c$. When $F=8$, the time mean RMSE tends to decrease until it reaches $c = 0.8$. After that, the time mean RMSE slowly rises. In both cases, all time mean RMSEs with $c > 0$ are lower than that of the ETKF ($c = 0$).

Similar results of the ETKF with I-ANA are observed in Fig.~\ref{fig:simL96_EnHF_nLoop20_ana_timeAvgRmse}. For both cases with $F=6$ and $F=8$, their time mean RMSE are monotonically decreasing functions with respect to $c$, and    all time mean RMSEs with $c > 0$ are lower than that of the ETKF ($c = 0$).

Fig.~\ref{fig:simL96_EnHF_nLoop20_im_timeAvgRmse} shows the time mean RMSEs of the ETKF with I-MTX. When $F=6$, the time mean RMSE decreases monotonically until it reaches $c = 0.4$. After that, the time mean RMSE rises rapidly. Moreover, filter divergence occurs for $c > 0.6$, possibly for the same reason as explained in the previous section. The result of $F=8$ is similar: the time mean RMSE decreases until $c=0.5$, and then increases as $c$ continues growing. Filter divergence also occurs when $c > 0.6$. Compared to the ETKF ($c=0$), the time mean RMSEs with $c>0$ are lower until $c$ reaches the turnaround point.

Through the above experiments, we have shown that, with suitable PLC values, the ETKFs of all three specific forms, namely, I-BG, I-ANA and I-MTX, exhibit relative robustness in comparison with the ETKF without any covariance inflation, which is consistent with the observations in the literature that an EnKF method with suitable covariance inflation may perform better than that without any covariance inflation (see, for example, \citealp{Hamill2009,VanLeeuwen2009} and the references therein). Different inflation schemes may result in different filter performance. For instance, the ETKF with I-ANA appears to have better performance than the other two schemes. The validity of this conclusion may depend on the system in assimilation, though, and may need to be verified case by case.

\section{Discussion and conclusion}

In this work we considered the applications of the KF and the HF to state estimation problem. We discussed the similarity and difference between the KF and the HF, and showed that the KF can be considered as a special case of the HF with the performance level equal to zero. For convenience of applying the $H_{\infty}$ filtering theory to sequential data assimilation, we introduced a variant, called the time-local HF, in which we suggested to solve the constraints in the HF locally (in time). Analogous to the EnKF methods, we proposed the ensemble version of the TLHF, called the ensemble time-local HF (EnTLHF), and showed that the EnTLHF can be constructed based on the EnKF. In addition, we established the connections of some specific forms of the EnTLHF to some EnKF methods equipped with certain covariance inflation techniques. 

Compared to existing works on covariance inflation in the EnKF, the $H_{\infty}$ filtering theory provides a theoretical framework that unifies various inflation techniques in the literature, and establishes the connection between covariance inflation and robustness. The $H_{\infty}$ filtering theory also provides an explicit definition of robustness and the associated mathematical description. Conceptually, this leads to the possibility to recast the problem of optimal covariance inflation as an optimization problem with multiple objectives, although further investigations will be needed for practical considerations. In addition, since the definition of robustness is filter-independent, the robustness property may be integrated into other types of nonlinear filters, e.g., the particle filter or the Gaussian sum filter \citep{Hoteit2008,Luo2008-spgsf1,vanLeeuwen-variance}, by imposing constraints similar to that in Eq.~(\ref{eq:local HF solution}). In our opinion, it might be less obvious to see how the above extensions can be made from the point of view of covariance inflation.

Through numerical experiments, we verified the relative robustness of three specific forms of the TLHF/EnTLHF in comparison with the KF/ETKF without covariance inflation. There are also some issues that have not been fully addressed in this work, for instance, the optimal choice of the performance level coefficient in conducting uncertainty inflation. Further investigations in these aspects will be considered in the future.

\section*{Acknowledgement}
We would like to thank two anonymous reviewers for their most constructive suggestions and comments that have significantly improved our work.

This publication is based on work supported by funds from the KAUST GCR Academic Excellence Alliance program.

\bibliographystyle{ametsoc}
\bibliography{references}

\begin{thebibliography}{39}
\providecommand{\natexlab}[1]{#1}
\providecommand{\url}[1]{\texttt{#1}}
\providecommand{\urlprefix}{URL }
\expandafter\ifx\csname urlstyle\endcsname\relax
  \providecommand{\doi}[1]{doi:\discretionary{}{}{}#1}\else
  \providecommand{\doi}{doi:\discretionary{}{}{}\begingroup
  \urlstyle{rm}\Url}\fi
\providecommand{\eprint}[2][]{\url{#2}}

\bibitem[{Anderson(2001)}]{Anderson-ensemble}
Anderson, J.~L., 2001: An ensemble adjustment {K}alman filter for data
  assimilation. \textit{Mon. Wea. Rev.}, \textbf{129}, 2884--2903.

\bibitem[{Anderson(2007)}]{Anderson2007}
Anderson, J.~L., 2007: An adaptive covariance inflation error correction
  algorithm for ensemble filters. \textit{Tellus}, \textbf{59A~(2)}, 210--224.

\bibitem[{Anderson(2009)}]{Anderson2009}
Anderson, J.~L., 2009: Spatially and temporally varying adaptive covariance
  inflation for ensemble filters. \textit{Tellus}, \textbf{61A}, 72--83.

\bibitem[{Anderson and Anderson(1999)}]{Anderson-Monte}
Anderson, J.~L. and S.~L. Anderson, 1999: A {M}onte {C}arlo implementation of
  the nonlinear filtering problem to produce ensemble assimilations and
  forecasts. \textit{Mon. Wea. Rev.}, \textbf{127}, 2741--2758.

\bibitem[{Beezley and Mandel(2007)}]{Beezley-morphing}
Beezley, J.~D. and J.~Mandel, 2007: Morphing ensemble {K}alman filters.
  \textit{Tellus}, \textbf{60A}, 131 -- 140.

\bibitem[{Bishop et~al.(2001)Bishop, Etherton, and Majumdar}]{Bishop-adaptive}
Bishop, C.~H., B.~J. Etherton, and S.~J. Majumdar, 2001: Adaptive sampling with
  ensemble transform {K}alman filter. {P}art {I}: theoretical aspects.
  \textit{Mon. Wea. Rev.}, \textbf{129}, 420--436.

\bibitem[{Burger(1985)}]{Burger1985}
Burger, J.~O., 1985: \textit{Statistical Decision Theory and {B}ayesian
  Analysis}. Springer-Verlag, 624 pp.

\bibitem[{Burgers et~al.(1998)Burgers, van Leeuwen, and
  Evensen}]{Burgers-analysis}
Burgers, G., P.~J. van Leeuwen, and G.~Evensen, 1998: On the analysis scheme in
  the ensemble {K}alman filter. \textit{Mon. Wea. Rev.}, \textbf{126},
  1719--1724.

\bibitem[{Cohn and Todling(1996)}]{Cohn1996-approx}
Cohn, S. and R.~Todling, 1996: Approximate data assimilation schemes for stable
  and unstable dynamics. \textit{J. Meteor. Soc. Japan}, \textbf{74}, 63--75.

\bibitem[{Evensen(1994)}]{Evensen-sequential}
Evensen, G., 1994: Sequential data assimilation with a nonlinear
  quasi-geostrophic model using {M}onte {C}arlo methods to forecast error
  statistics. \textit{J. Geophys. Res.}, \textbf{99(C5)}, 10\,143--10\,162.

\bibitem[{Evensen(2003)}]{Evensen-ensemble}
Evensen, G., 2003: The ensemble {K}alman filter: theoretical formulation and
  practical implementation. \textit{Ocean Dyn.}, \textbf{53}, 343--367.

\bibitem[{Evensen and van Leeuwen(1996)}]{Evensen-assimilation}
Evensen, G. and P.~J. van Leeuwen, 1996: Assimilation of geosat altimeter data
  for the aghulas current using the ensemble {K}alman filter with a
  quasi-geostrophic model. \textit{Mon. Wea. Rev.}, \textbf{124}, 85--96.

\bibitem[{Francis(1987)}]{Francis1987}
Francis, B.~A., 1987: \textit{A Course in $\text{H}_{\infty}$ Control Theory}.
  Springer-Verlag, 156 pp.

\bibitem[{Hamill et~al.(2009)Hamill, Whitaker, Anderson, and
  Snyder}]{Hamill2009}
Hamill, T.~M., J.~S. Whitaker, J.~L. Anderson, and C.~Snyder, 2009: Comments on
  ``{S}igma-point {K}alman filter data assimilation methods for strongly
  nonlinear systems''. \textit{Journal of the Atmospheric Sciences},
  \textbf{66}, 3498--3500.

\bibitem[{Hamill et~al.(2001)Hamill, Whitaker, and Snyder}]{Hamill-distance}
Hamill, T.~M., J.~S. Whitaker, and C.~Snyder, 2001: Distance-dependent
  filtering of background error covariance estimates in an ensemble {K}alman
  filter. \textit{Mon. Wea. Rev.}, \textbf{129}, 2776--2790.

\bibitem[{Hoteit and Pham(2004)}]{Hoteit2004-adaptively}
Hoteit, I. and D.~T. Pham, 2004: An adaptively reduced-order extended kalman
  filter for data assimilation in the tropical pacific. \textit{Journal of
  Marine Systems}, \textbf{45~(3-4)}, 173--188.

\bibitem[{Hoteit et~al.(2001)Hoteit, Pham, and Blum}]{Hoteit2001}
Hoteit, I., D.~T. Pham, and J.~Blum, 2001: A semi-evolutive partially local
  filer for data assimilation. \textit{Marine Pollution Bulletin}, \textbf{43},
  164--174.

\bibitem[{Hoteit et~al.(2002)Hoteit, Pham, and Blum}]{Hoteit2002}
Hoteit, I., D.~T. Pham, and J.~Blum, 2002: A simplified reduced order {K}alman
  filtering and application to altimetric data assimilation in {T}ropical
  {P}acific. \textit{Journal of Marine Systems}, \textbf{36}, 101--127.

\bibitem[{Hoteit et~al.(2008)Hoteit, Pham, Triantafyllou, and
  Korres}]{Hoteit2008}
Hoteit, I., D.~T. Pham, G.~Triantafyllou, and G.~Korres, 2008: A new
  approximate solution of the optimal nonlinear filter for data assimilation in
  meteorology and oceanography. \textit{Mon. Wea. Rev.}, \textbf{136},
  317--334.

\bibitem[{Houtekamer and Mitchell(1998)}]{Houtekamer1998}
Houtekamer, P.~L. and H.~L. Mitchell, 1998: Data assimilation using an ensemble
  {K}alman filter technique. \textit{Mon. Wea. Rev.}, \textbf{126}, 796--811.

\bibitem[{Jazwinski(1970)}]{Jazwinski1970}
Jazwinski, A.~H., 1970: \textit{Stochastic Processes and Filtering Theory}.
  Academic Press.

\bibitem[{Kalman(1960)}]{Kalman-new}
Kalman, R., 1960: A new approach to linear filtering and prediction problems.
  \textit{Trans. ASME, Ser. D, J. Basic Eng.}, \textbf{82}, 35--45.

\bibitem[{Lorenz(1996)}]{Lorenz-predictability}
Lorenz, E.~N., 1996: Predictability-a problem partly solved.
  \textit{Predictability}, T.~Palmer, Ed., ECMWF, Reading, UK.

\bibitem[{Lorenz and Emanuel(1998)}]{Lorenz-optimal}
Lorenz, E.~N. and K.~A. Emanuel, 1998: Optimal sites for supplementary weather
  observations: Simulation with a small model. \textit{J. Atmos. Sci.},
  \textbf{55}, 399--414.

\bibitem[{Luo and Moroz(2009)}]{Luo-ensemble}
Luo, X. and I.~M. Moroz, 2009: Ensemble {K}alman filter with the unscented
  transform. \textit{Physica D}, \textbf{238}, 549--562.

\bibitem[{Luo et~al.(2010)Luo, Moroz, and Hoteit}]{Luo2008-spgsf1}
Luo, X., I.~M. Moroz, and I.~Hoteit, 2010: Scaled unscented transform
  {G}aussian sum filter: Theory and application. \textit{Physica D},
  \textbf{239}, 684--701.

\bibitem[{Nerger et~al.(2005)Nerger, Hiller, and Schr\"{o}ter}]{Nerger2005}
Nerger, L., L.~Hiller, and J.~Schr\"{o}ter, 2005: A comparison of error
  subspace {K}alman filters. \textit{Tellus}, \textbf{57A}, 715--735.

\bibitem[{Ott et~al.(2004)}]{Ott-local}
Ott, E., et~al., 2004: A local ensemble {K}alman filter for atmospheric data
  assimilation. \textit{Tellus}, \textbf{56A}, 415--428.

\bibitem[{Pham et~al.(1998)Pham, Verron, and Roubaud}]{Pham1998}
Pham, D.~T., J.~Verron, and M.~C. Roubaud, 1998: A singular evolutive extended
  {K}alman filter for data assimilation in oceanography. \textit{Journal of
  Marine Systems}, \textbf{16}, 323--340.

\bibitem[{Schlee et~al.(1967)Schlee, Standish, and Toda}]{Schlee1967}
Schlee, F.~H., C.~J. Standish, and N.~F. Toda, 1967: Divergence in the kalman
  filter. \textit{AIAA Journal}, \textbf{5}, 1114--1120.

\bibitem[{Shaked and Berman(1995)}]{Shaked1995}
Shaked, U. and N.~Berman, 1995: ${H}_{\infty}$ nonlinear filtering of
  discrete-time processes. \textit{IEEE Transactions on Signal Processing},
  \textbf{43~(9)}, 2205--2209.

\bibitem[{Simon(2006)}]{Simon2006}
Simon, D., 2006: \textit{Optimal State Estimation: {K}alman, {H}-Infinity, and
  Nonlinear Approaches}. Wiley-Interscience, 552 pp.

\bibitem[{Tippett et~al.(2003)Tippett, Anderson, Bishop, Hamill, and
  Whitaker}]{Tippett-ensemble}
Tippett, M.~K., J.~L. Anderson, C.~H. Bishop, T.~M. Hamill, and J.~S. Whitaker,
  2003: Ensemble square root filters. \textit{Mon. Wea. Rev.}, \textbf{131},
  1485--1490.

\bibitem[{Van~Leeuwen(2003)}]{vanLeeuwen-variance}
Van~Leeuwen, P.~J., 2003: A variance minimizing filter for large-scale
  applications. \textit{Mon. Wea. Rev.}, \textbf{131}, 2071--2084.

\bibitem[{Van~Leeuwen(2009)}]{VanLeeuwen2009}
Van~Leeuwen, P.~J., 2009: Particle filtering in geophysical systems.
  \textit{Mon. Wea. Rev.}, \textbf{137}, 4089--4114.

\bibitem[{Verlaan and Heemink(1997)}]{Verlaan1997}
Verlaan, M. and A.~W. Heemink, 1997: Tidal flow forecasting using reduced rank
  square root filters. \textit{Stochastic Hydrology and Hydraulics},
  \textbf{11}, 349--368.

\bibitem[{Wang and Cai(2008)}]{Wang2008455}
Wang, D. and X.~Cai, 2008: Robust data assimilation in hydrological modeling -
  a comparison of {K}alman and {H}-infinity filters. \textit{Advances in Water
  Resources}, \textbf{31~(3)}, 455 -- 472.

\bibitem[{Whitaker and Hamill(2002)}]{Whitaker-ensemble}
Whitaker, J.~S. and T.~M. Hamill, 2002: Ensemble data assimilation without
  perturbed observations. \textit{Mon. Wea. Rev.}, \textbf{130}, 1913--1924.

\bibitem[{Zupanski(2005)}]{Zupanski-maximum}
Zupanski, M., 2005: Maximum likelihood ensemble filter: theoretical aspects.
  \textit{Mon. Wea. Rev.}, \textbf{133}, 1710--1726.

\end{thebibliography}

\renewcommand{\thefigure}{\arabic{figure}}
\renewcommand{\thetable}{\arabic{table}}
\clearpage


\clearpage
\begin{figure*} 
\centering
\includegraphics[width=\textwidth]{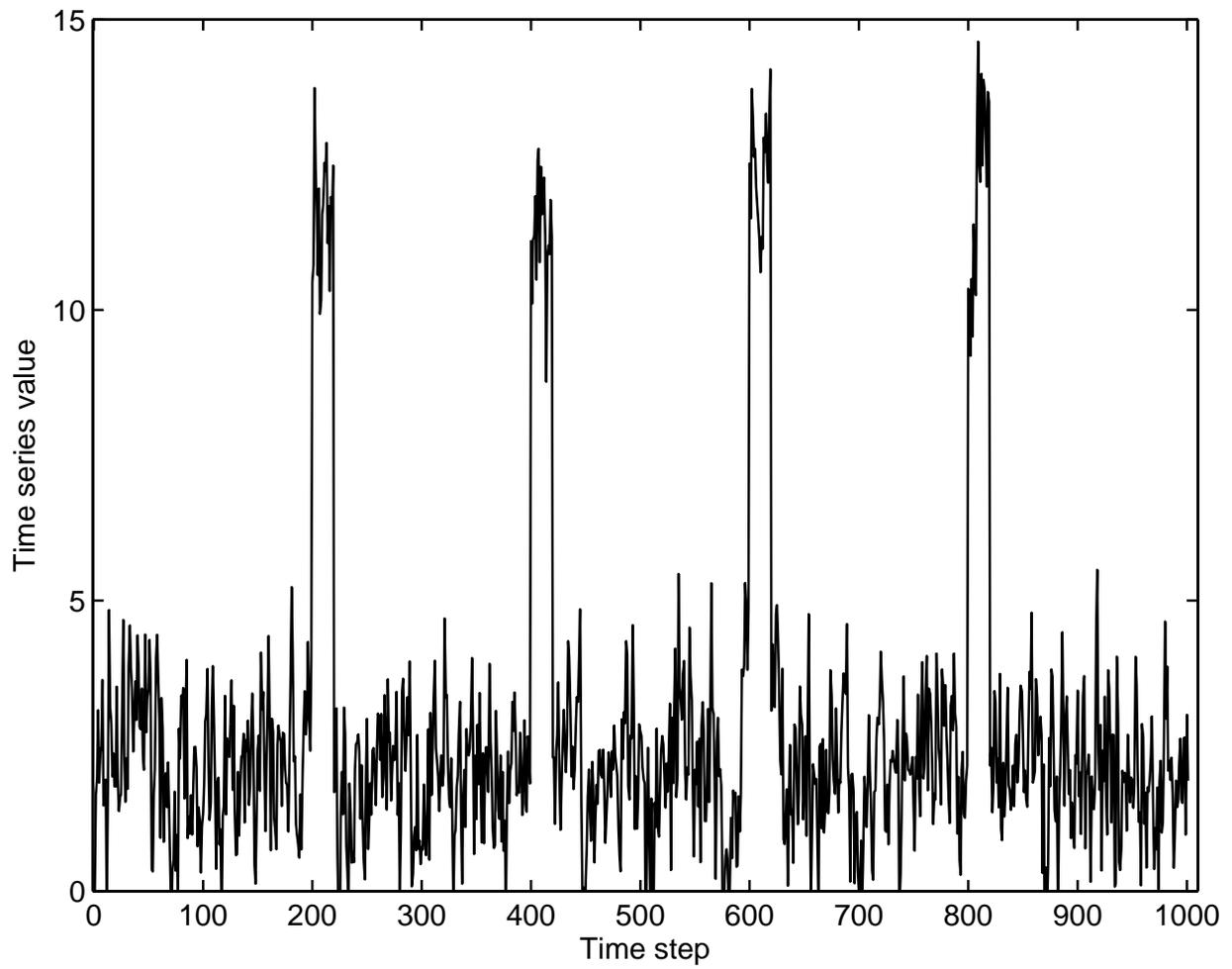}
\caption{\label{fig:simAR1_time_series} A time series generated by Eq.~(\ref{eq:true_one_dimensional_regression_model}).}
\end{figure*}


\clearpage
\begin{figure*} 
\centering
\includegraphics[width=\textwidth]{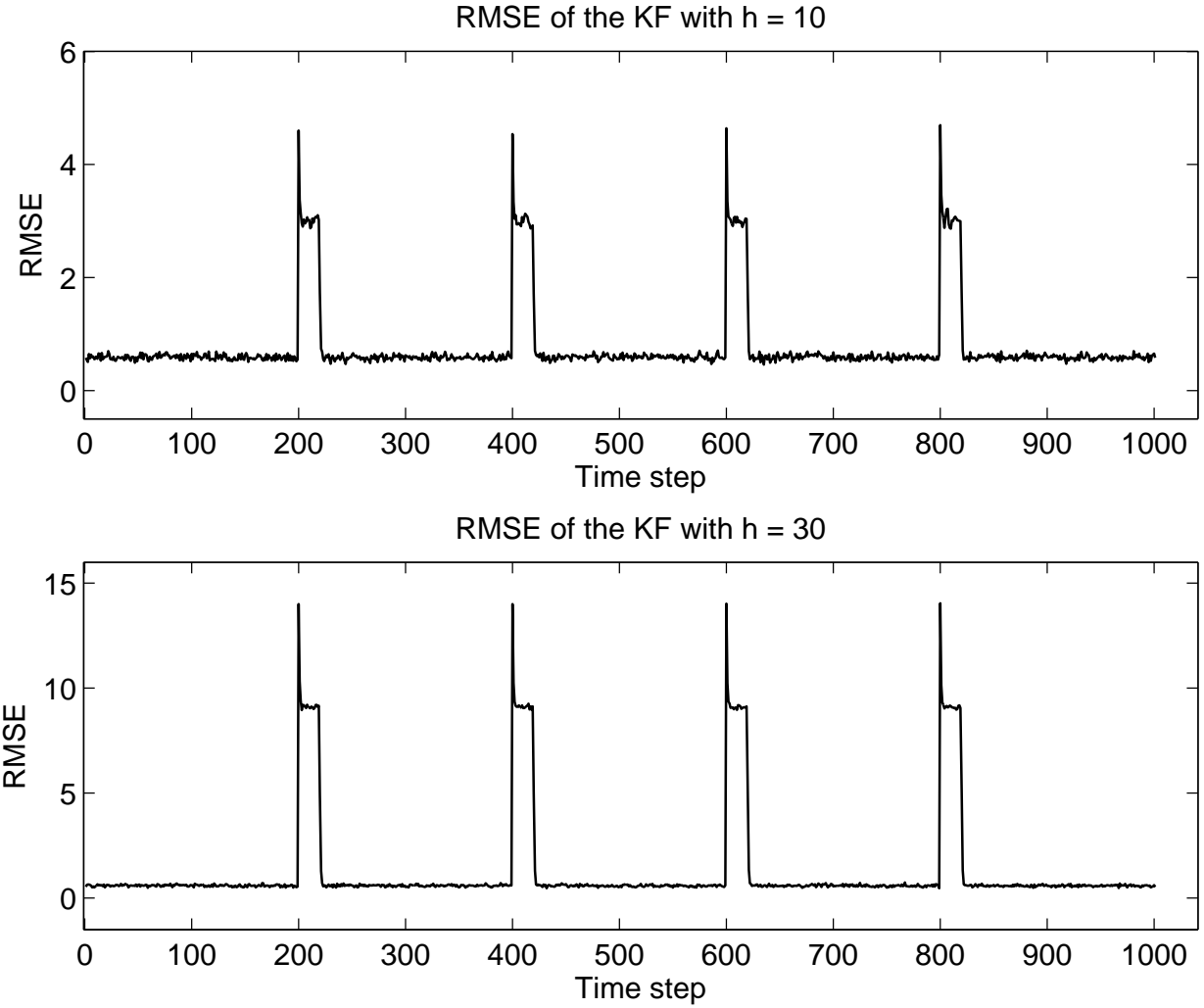}
\caption{ \label{fig:simAR1_KF_varying_h1030} RMSE of the KF in assimilating the regression model. The jump heights $h$ are $10$ (upper panel) and $30$ (lower panel), respectively.}
\end{figure*}

\clearpage
\begin{figure*} 
\centering
\includegraphics[width=\textwidth]{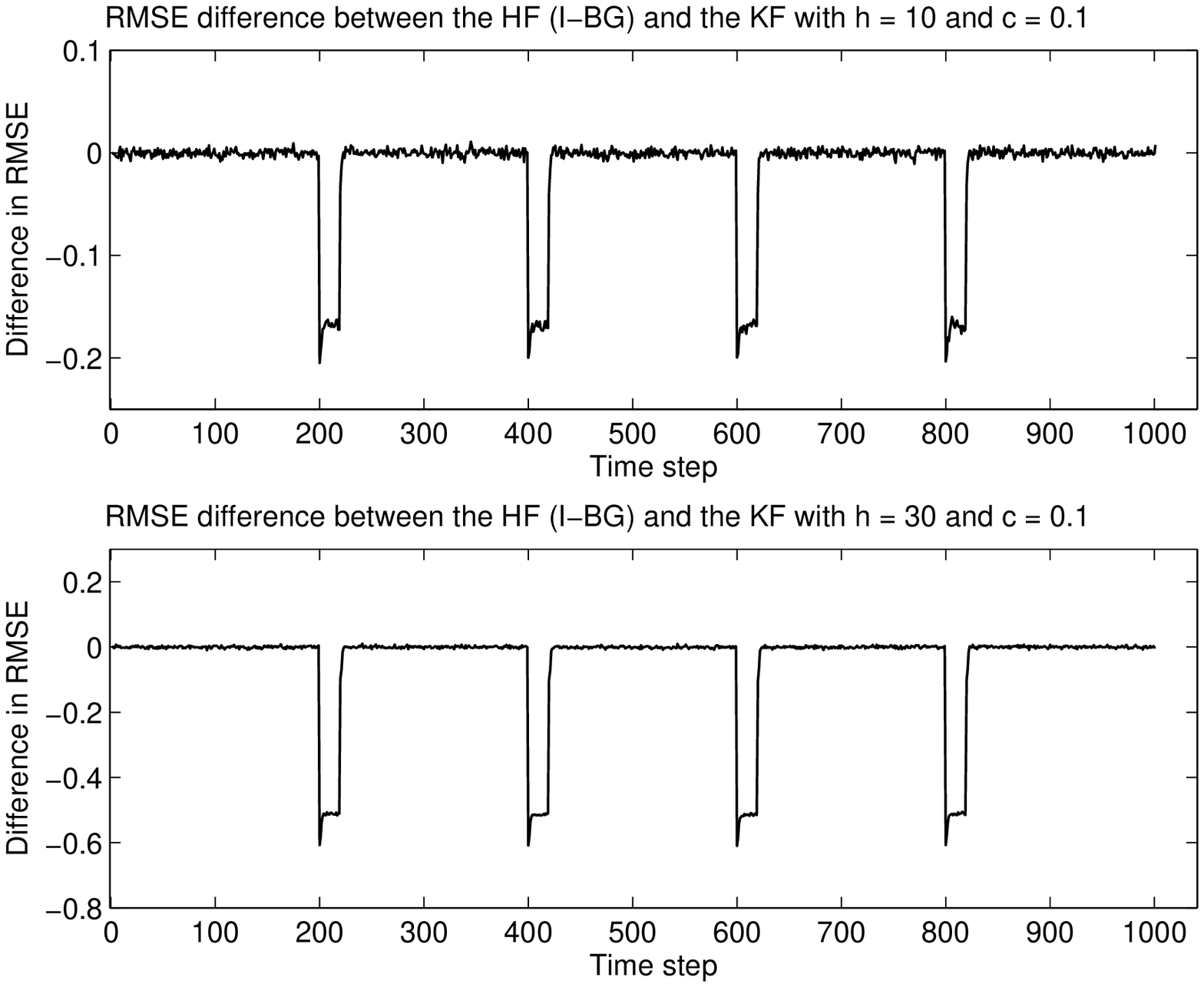}
\caption{ \label{fig:simAR1_HF_KF_varying_h1030_bg_c01} RMSE difference between the TLHF of I-BG and the KF in assimilating the regression model. The jump heights $h$ are $10$ (upper panel) and $30$ (lower panel), respectively. In the TLHF, the PLC $c=0.1$.}
\end{figure*}

\clearpage
\begin{figure*} 
\centering
\includegraphics[width=\textwidth]{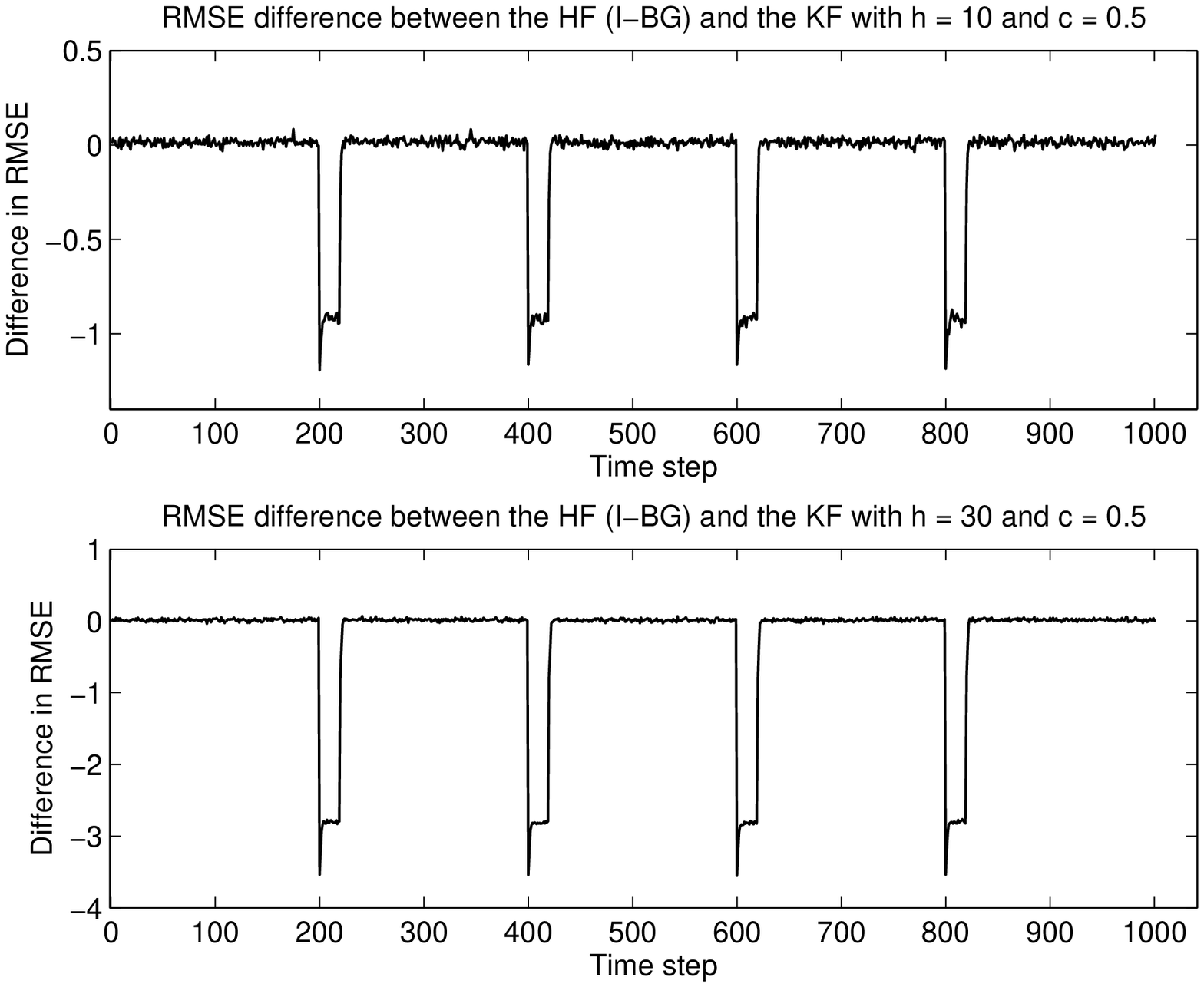}
\caption{ \label{fig:simAR1_HF_KF_varying_h1030_bg_c05} RMSE difference between the TLHF of I-BG and the KF in assimilating the regression model. The jump heights $h$ are $10$ (upper panel) and $30$ (lower panel), respectively. In the TLHF, the PLC $c=0.5$.}
\end{figure*}

\clearpage
\begin{figure*} 
\centering
\includegraphics[width=\textwidth]{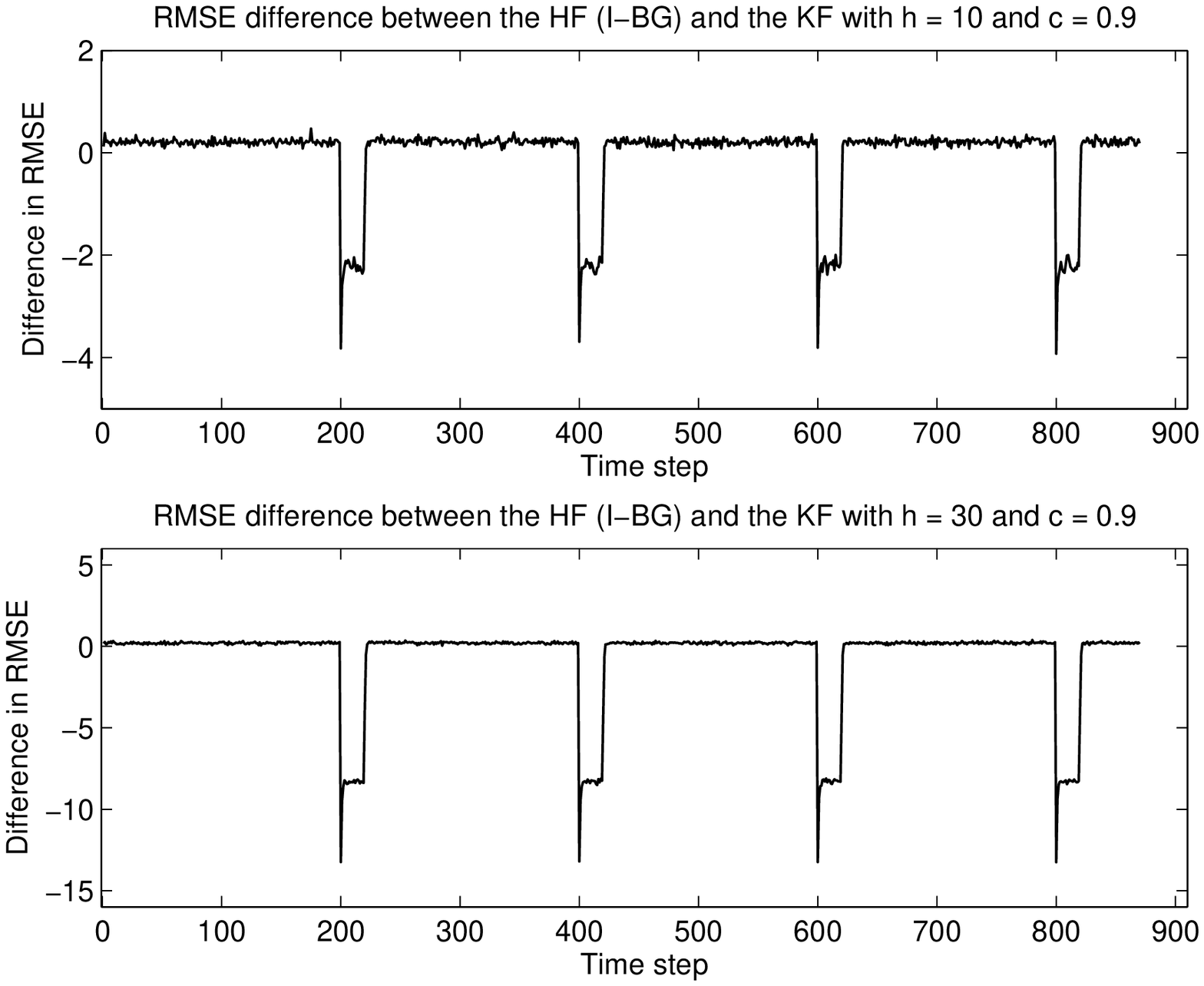}
\caption{ \label{fig:simAR1_HF_KF_varying_h1030_bg_c09} RMSE difference between the TLHF of I-BG and the KF in assimilating the regression model. The jump heights $h$ are $10$ (upper panel) and $30$ (lower panel), respectively. In the TLHF, the PLC $c=0.9$.}
\end{figure*}

\clearpage
\begin{figure*} 
\centering
\includegraphics[width=\textwidth]{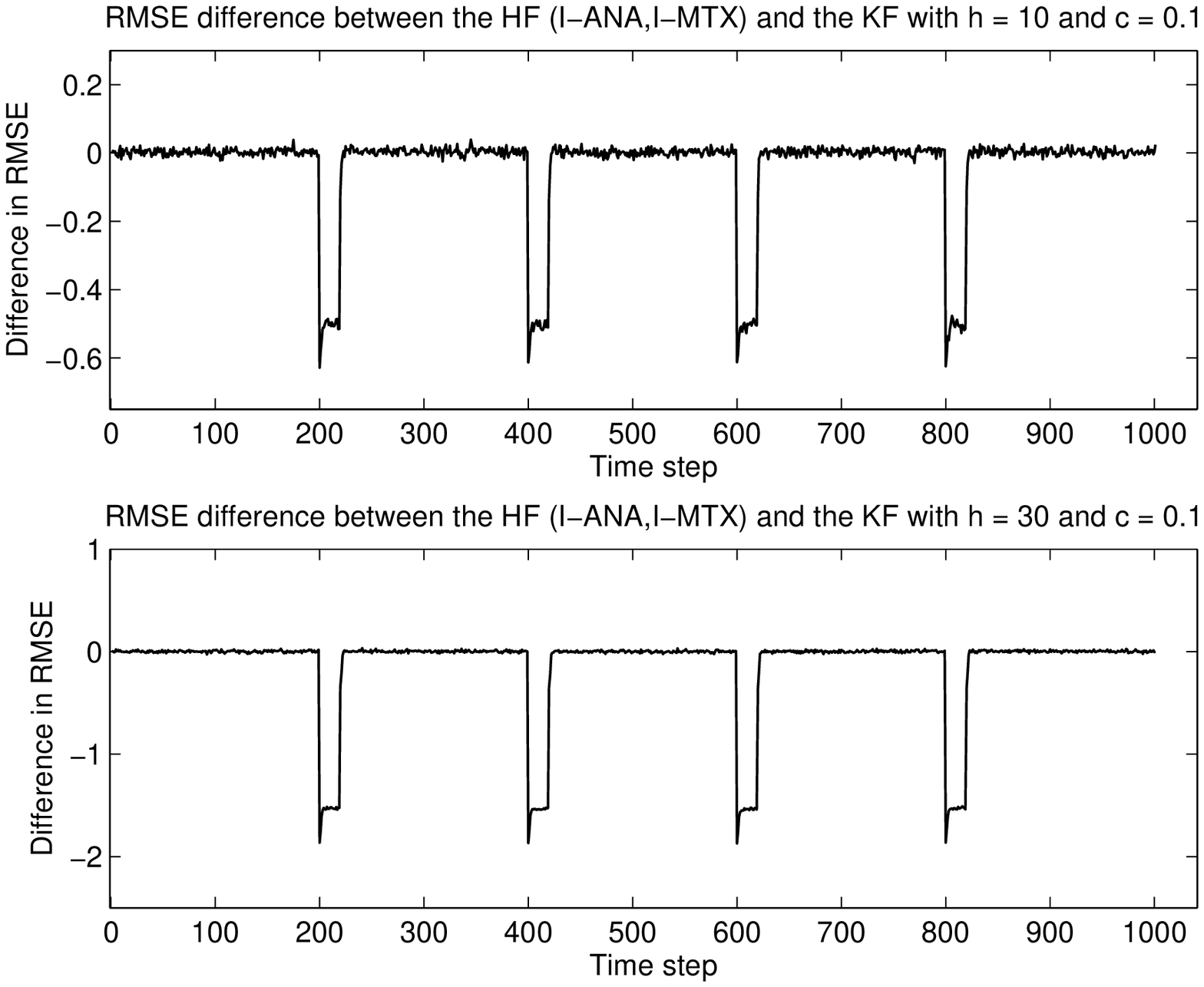}
\caption{ \label{fig:simAR1_HF_KF_varying_h1030_ana_c01} RMSE difference between the TLHF of I-ANA and I-MTX and the KF in assimilating the regression model. The jump heights $h$ are $10$ (upper panel) and $30$ (lower panel), respectively. In the TLHF, the PLC $c=0.1$.}
\end{figure*}

\clearpage
\begin{figure*} 
\centering
\includegraphics[width=\textwidth]{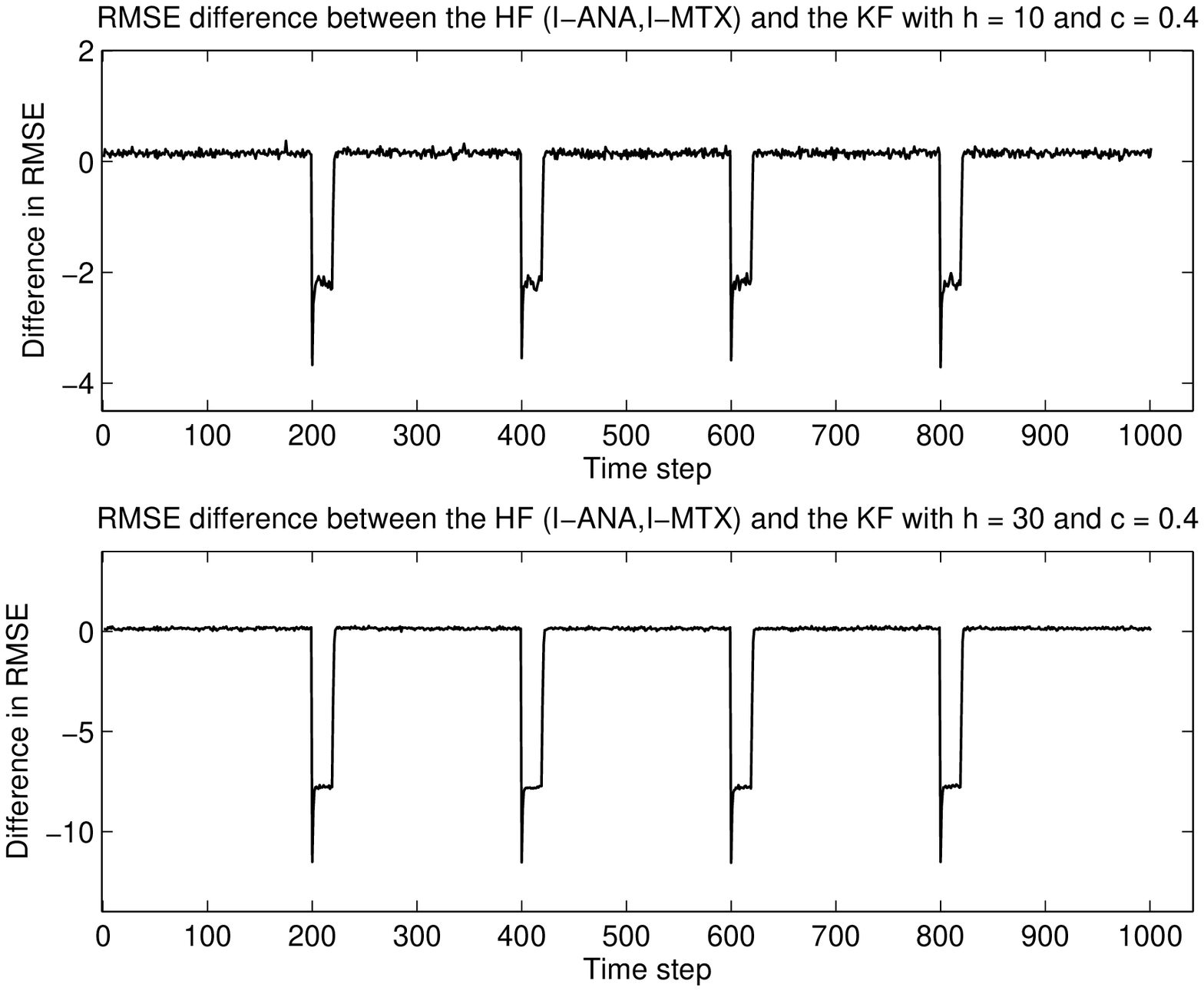}
\caption{ \label{fig:simAR1_HF_KF_varying_h1030_ana_c04} RMSE difference between the TLHF of I-ANA and I-MTX and the KF in assimilating the regression model. The jump heights $h$ are $10$ (upper panel) and $30$ (lower panel), respectively. In the TLHF, the PLC $c=0.4$.}
\end{figure*}

\clearpage
\begin{figure*} 
\centering
\includegraphics[width=\textwidth]{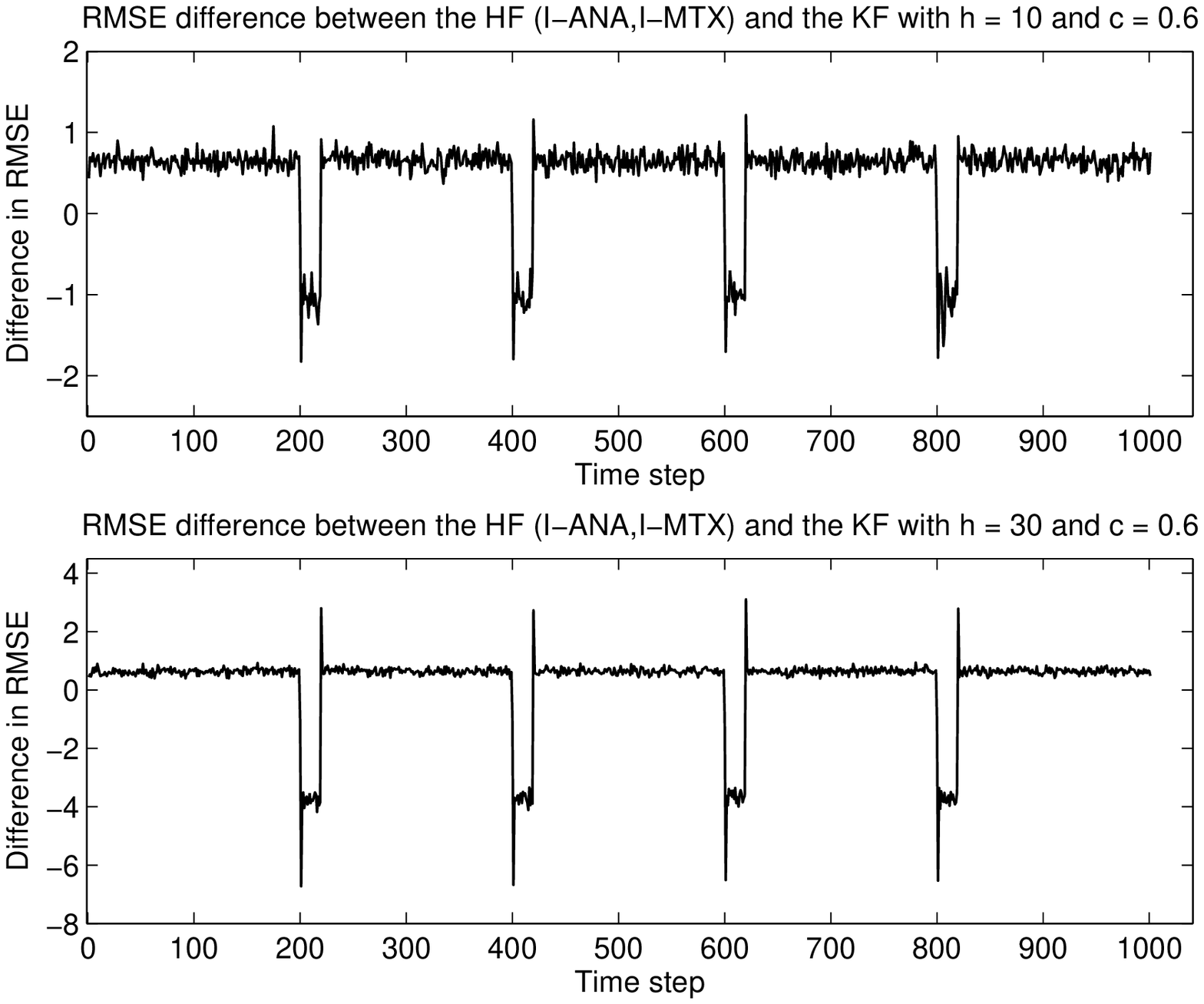}
\caption{ \label{fig:simAR1_HF_KF_varying_h1030_ana_c06} RMSE difference between the TLHF of I-ANA and I-MTX and the KF in assimilating the regression model. The jump heights $h$ are $10$ (upper panel) and $30$ (lower panel), respectively. In the TLHF, the PLC $c=0.6$.}
\end{figure*}
\clearpage
\begin{figure*} 
\centering
\includegraphics[width=\textwidth]{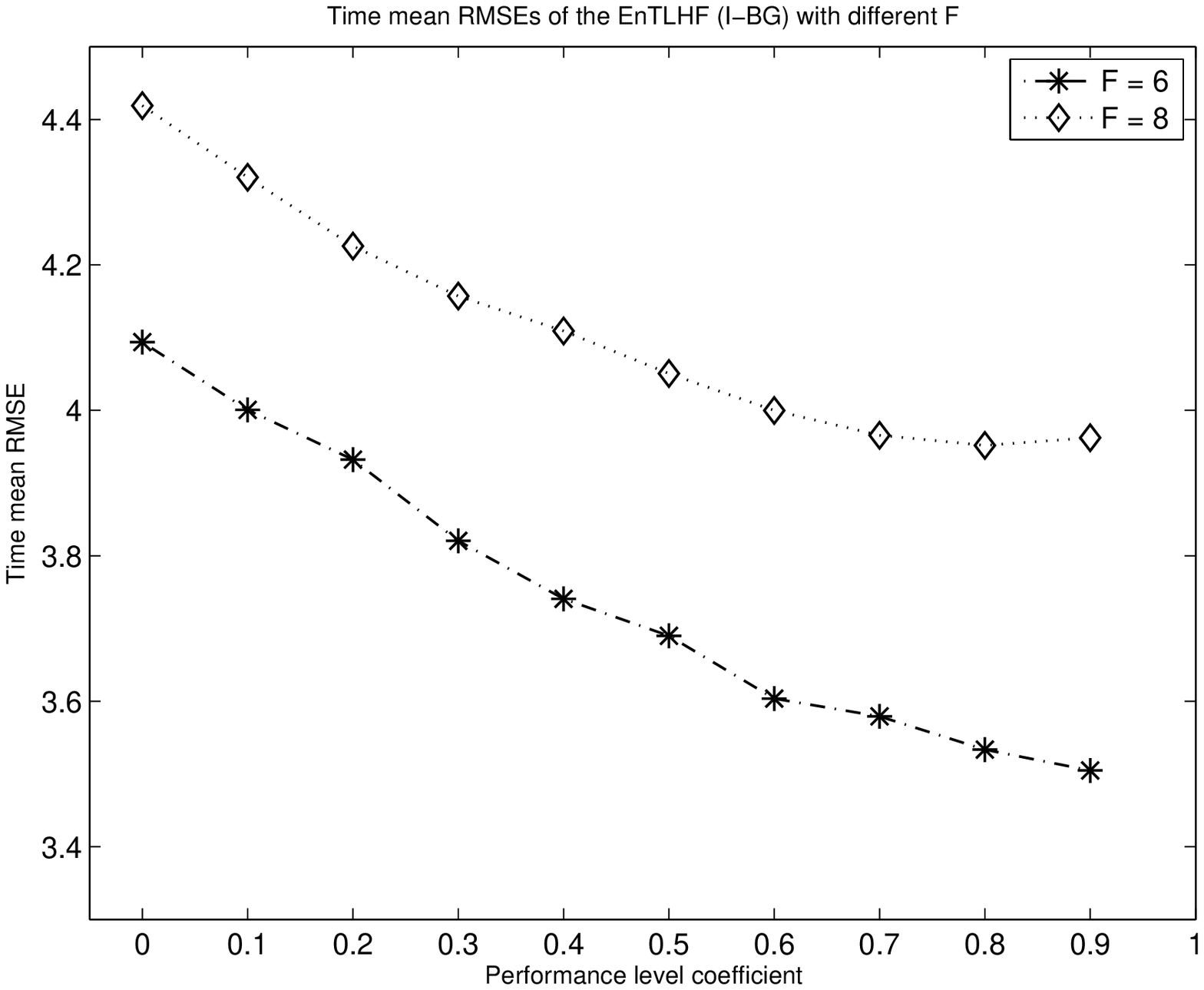}
\caption{ \label{fig:simL96_EnHF_nLoop20_bg_timeAvgRmse} Time mean RMSE of the EnTLHF of I-BG as a function of the PLC in assimilating the L96 model. The values of the parameter $F$ are $6$ (dash-dotted) and $8$ (dotted), respectively. The EnTLHF reduces to the ETKF when the PLC $c=0$.}
\end{figure*}

\clearpage
\begin{figure*} 
\centering
\includegraphics[width=\textwidth]{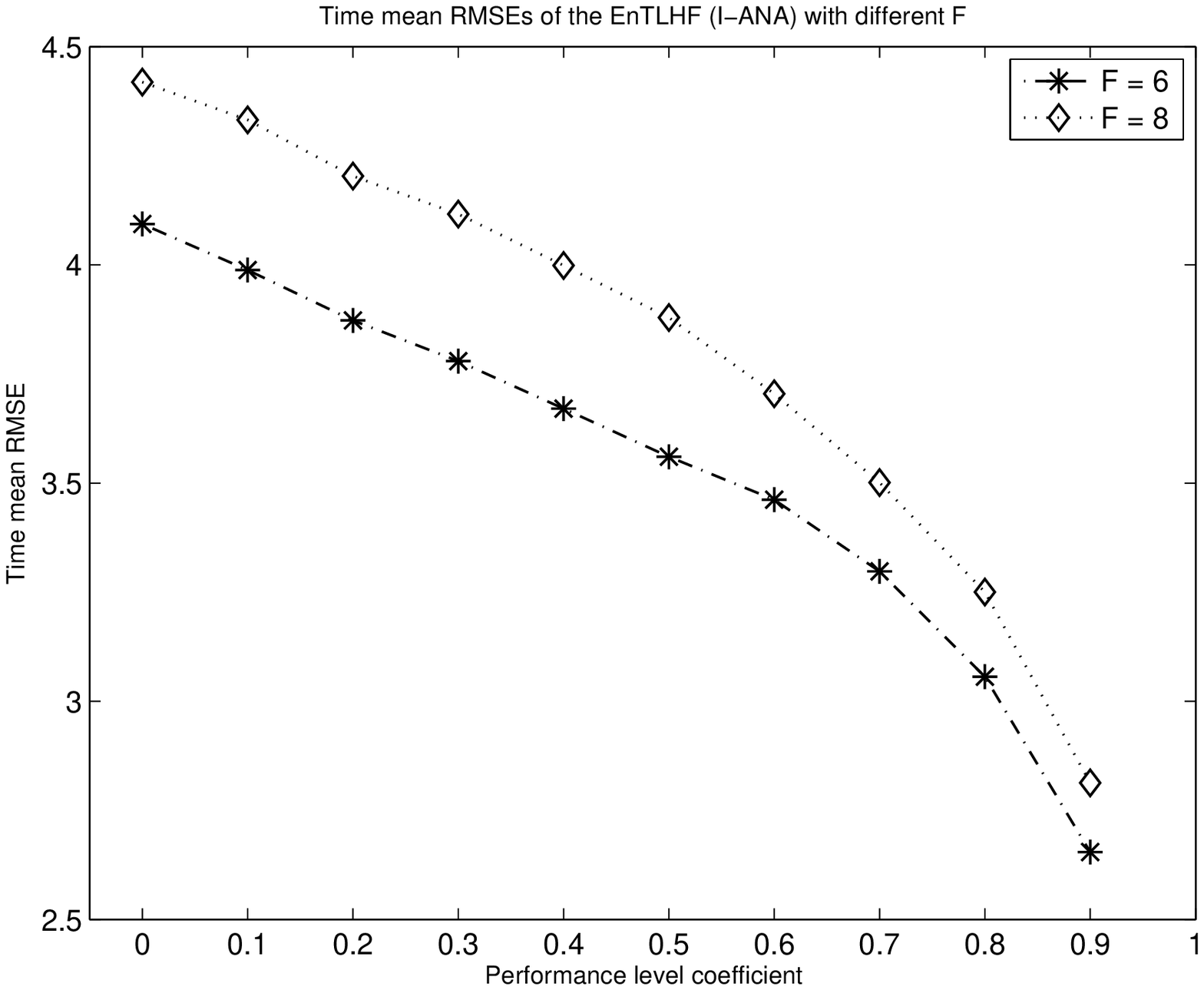}
\caption{ \label{fig:simL96_EnHF_nLoop20_ana_timeAvgRmse} Time mean RMSE of the EnTLHF of I-ANA as a function of the PLC in assimilating the L96 model. The values of the parameter $F$ are $6$ (dash-dotted) and $8$ (dotted), respectively. The EnTLHF reduces to the ETKF when the PLC $c=0$.}
\end{figure*}

\clearpage
\begin{figure*} 
\centering
\includegraphics[width=\textwidth]{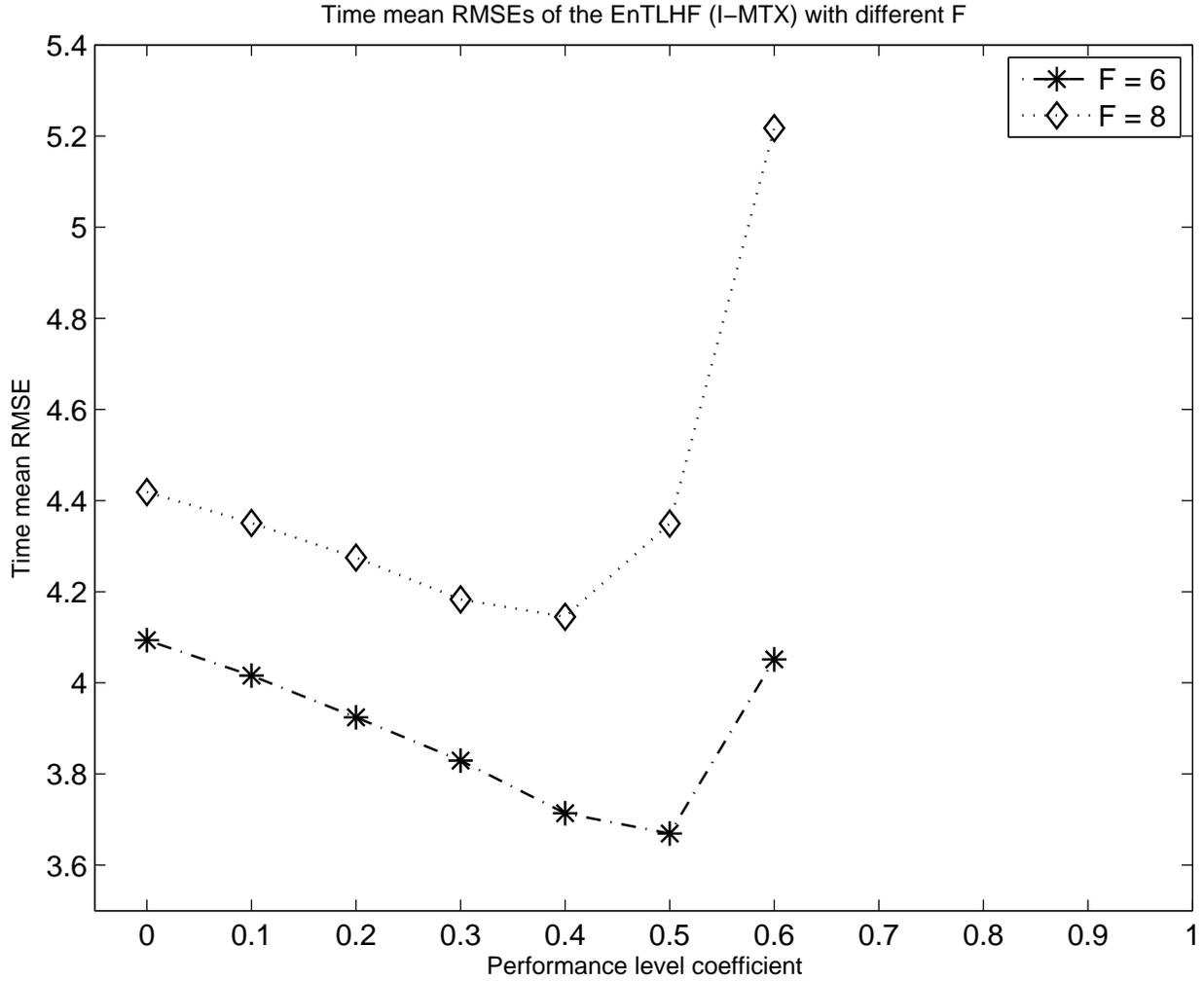}
\caption{ \label{fig:simL96_EnHF_nLoop20_im_timeAvgRmse} Time mean RMSE of the EnTLHF of I-MTX as a function of the PLC in assimilating the L96 model. The values of the parameter $F$ are $6$ (dash-dotted) and $8$ (dotted), respectively. The EnTLHF reduces to the ETKF when the PLC $c=0$. Filter divergence occurs for $c > 0.6$.}
\end{figure*}
\end{document}